\documentclass[aps,twocolumn,superscriptaddress,floatfix]{revtex4}
\usepackage[]{graphicx}
\usepackage{amsmath,amssymb,amsfonts}
\usepackage[usenames,dvipsnames]{color}
\usepackage{soul}
\usepackage{bm}
\usepackage{mathtools}

\begin{document}


\title{Towards Topological Quasi-Freestanding Stanene via Substrate Engineering}

\author{Domenico Di Sante}
\affiliation{Institut f\"{u}r Theoretische Physik und Astrophysik, Universit\"{a}t W\"{u}rzburg, Am Hubland Campus S\"{u}d, W\"{u}rzburg 97074, Germany}\email{domenico.disante@physik.uni-wuerzburg.de}

\author{Philipp Eck}
\affiliation{Institut f\"{u}r Theoretische Physik und Astrophysik, Universit\"{a}t W\"{u}rzburg, Am Hubland Campus S\"{u}d, W\"{u}rzburg 97074, Germany}

\author{Maximilian Bauernfeind}
\affiliation{Physikalisches Institut and R\"{o}ntgen Research Center for Complex Material Systems, Universit\"{a}t W\"{u}rzburg, Am Hubland Campus S\"{u}d, W\"{u}rzburg 97074, Germany}

\author{Marius Will}
\affiliation{Physikalisches Institut and R\"{o}ntgen Research Center for Complex Material Systems, Universit\"{a}t W\"{u}rzburg, Am Hubland Campus S\"{u}d, W\"{u}rzburg 97074, Germany}

\author{Ronny Thomale}
\affiliation{Institut f\"{u}r Theoretische Physik und Astrophysik, Universit\"{a}t W\"{u}rzburg, Am Hubland Campus S\"{u}d, W\"{u}rzburg 97074, Germany}

\author{J\"{o}rg Sch\"{a}fer}
\affiliation{Physikalisches Institut and R\"{o}ntgen Research Center for Complex Material Systems, Universit\"{a}t W\"{u}rzburg, Am Hubland Campus S\"{u}d, W\"{u}rzburg 97074, Germany}

\author{Ralph Claessen}
\affiliation{Physikalisches Institut and R\"{o}ntgen Research Center for Complex Material Systems, Universit\"{a}t W\"{u}rzburg, Am Hubland Campus S\"{u}d, W\"{u}rzburg 97074, Germany}

\author{Giorgio Sangiovanni}
\affiliation{Institut f\"{u}r Theoretische Physik und Astrophysik, Universit\"{a}t W\"{u}rzburg, Am Hubland Campus S\"{u}d, W\"{u}rzburg 97074, Germany}

\date{\today}

\begin{abstract}
In search for a new generation of spintronics hardware, material
candidates for room temperature quantum spin Hall effect (QSHE) have
become a contemporary focus of investigation. Inspired by the original
proposal for QSHE in graphene, several heterostructures have been
synthesized, aiming at a hexagonal monolayer of heavier group IV
elements promoting the QSHE bulk gap via increased spin-orbit
coupling. So far, the monolayer/substrate coupling, which can
manifest itself in strain, deformation, and hybridization, has proven to
be detrimental to the aspired QSHE conditions for the monolayer.
For stanene, the Sn analogue of graphene, we
investigate how an interposing buffer layer mediates between monolayer
and substrate in order to optimize the QSHE setting. From a detailed
density functional theory study, we highlight the principal mechanisms
induced by such a buffer layer to accomplish quasi-freestanding stanene
in its QSHE phase. We complement our theoretical predictions by
presenting the first attempts to grow a buffer layer on SiC(0001)
on which stanene can be deposited.
\end{abstract}

\maketitle


\section{Introduction}
Quantum spin Hall (QSH) systems are 
two-dimensional bulk non-conducting materials featuring topologically
non-trivial conducting edge
modes, whose stability against external perturbations is ensured by
time reversal symmetry \cite{RPMHasan,RPMQi}. After the first theoretical prediction in
graphene by Kane and Mele \cite{KM1,KM2} and the subsequent prediction
and experimental
realization in HgTe/CdTe quantum wells \cite{Bernevig,Konig}, it became
evident that the small bulk energy gap represents the hardest
obstacle to render QSH materials operative at room
temperature and, as such, relevant to technological applications.

An initial plausible way of increasing the bulk gap is to enhance the
spin-orbit coupling strength of the constituting atoms.
Within the group-IV elements, freestanding honeycomb-like structures made by Si, Ge,
and Sn atoms, dubbed silicene, germanene, and stanene, respectively, were
estimated in theory to exhibit gaps up to 100 meV
\cite{2DXenes,Stanene,GeisslerNJP}. A realistic implementation, however,
always requires the stabilization of freestanding 2D layers on a
supporting substrate, with no guarantee that the resulting symmetry
breaking keeps the QSH phase intact. For instance, this detrimental effect to the
QSHE phase is seen
when an insulating MoS$_2$ substrate stabilizes the growth of
germanene \cite{GermaneneMoS2} or a Bi$_2$Te$_3$ template is used to accommodate stanene
flakes \cite{Zhu2015}. In both cases, the topological bulk gap does not survive
the interaction with the substrate, and either yields a metallic or
trivially insulating monolayer.

By contrast, a constructive effect of the substrate is observed for bismuthene on SiC \cite{Reis287}. There,
the strong hybridization between the substrate and the Bi $\pi$ orbitals
leaves a sizable topological bulk gap of 0.8 eV, highlighting bismuthene as the first
material realization of a QSH system operable at ambient temperature.
Monolayer WTe$_2$  is another recent example where the interaction between
the 2D layer and the substrate stabilizes the QSH phase by opening a finite
gap, thus avoiding contributions from the bulk material which may
be detrimental to the edge conductivity \cite{WTe2_1,WTe2_2,WTe2}.

On the basis of these observations, it is evident that any meaningful theoretical investigation
must properly take the role of substrates  into consideration. With a
particular focus
on stanene, several studies have attempted to include these effects \cite{Zhu2015,StaneneInSb,ZeyuanNi,Bechstedt,Matusalem}, while
only a few of them have considered technologically achievable, and hence relevant, substrates
\cite{Bechstedt,Matusalem}. Rather, our objective is to address
commercial wafers of wide-gap semiconductors as possible substrate
candidates, in order to follow the motif of reaching proposals that
might allow for  the integration of QSH physics into ambient
life devices. SiC(0001) turns out to
be a promising way along this direction \cite{PRLJoerg}. However, dangling
bond passivation is a well-known problem of semiconductor surfaces. The
presence of highly reactive surface charge may be
detrimental to the QSH phase. Recent theoretical investigations proceeded by removing
the effect of the dangling bonds via hydrogen (H) saturation \cite{Bechstedt,Glass2016}.
Despite its modeling efficiency, this cannot be considered a
viable solution, since no evidence of a 2D material grown on a pure
H-saturated surface has so far been reported in the literature.

For silicene and germanene, several studies report on attempts of reducing the interaction with the substrate \cite{houssa2010,kanno2014,dacapito2016,niNanoLett12,dueSciAdv2016,kaloni2013}.
Regarding carbon, one monolayer deposited on SiC does not form graphene, as the symmetries and electronic properties are destroyed by the strong binding. Moving towards almost freestanding graphene, one can either grow further carbon on this first layer, which in fact takes care of saturating the dangling bonds, or intercalate ``spacers'' such as H, Au or O between the first layer and the substrate \cite{virojanadaraPRB82,enderleinNJP12,oidaPRB82,varchonPRL99,watcharinyanonSurfSci2011,Emtsev2009,Zhou2007}. 
Here, we explore another possibility, namely realizing the QSH phase in stanene
on SiC(0001) by the insertion of a buffer layer of group-III and group-V
elements. 
This way, the valence electrons from the buffer atoms saturate the SiC dangling bonds, keeping at the same time a favorable lattice matching.
The positive effect of the buffer is counterbalanced by an unavoidable
monolayer/substrate hybridization which, as we will explicate
in the following, turns out not to damage the QSHE under particular
conditions, and instead to even structurally stabilize stanene.

For buckled geometries, as it is also the case for stanene, a concomitant
staggered potential (Semenoff mass \cite{Semenoff}) emerges, which for
specific buffers is strong enough to prevent the QSH formation~\cite{KM1}. 
Via a proper choice of the buffer characteristics, we tune the quantitative impact of
the Semenoff mass term.
While our DFT calculations show that on Al, Ga, In, and Tl, stanene is
topologically trivial, we successfully minimize the detrimental effect
of the Semenoff mass through switching to buffers made of group-V elements.  
Our calculations indeed demonstrate that the topology in
stanene can be comfortably stabilized for buffer layers made of P and
As atoms, which might stimulate further experimental
progress along this direction.


\section{Methods}
For our theoretical study of stanene on buffered SiC(0001) we employ
state-of-the-art first-principles calculations based on density
functional theory as implemented in the Vienna ab-initio simulation
package (VASP) \cite{VASP1}, within the projector-augmented-plane-wave (PAW)
method \cite{VASP2,PAW}. The generalized gradient approximation as parametrized by
the PBE-GGA functional for the exchange-correlation potential is used
\cite{PBE}, as well as different flavours of revised and van der Waals corrected functionals \cite{PBEsol,VDW1,VDW2,VDWopt},
by expanding the Kohn-Sham wavefunctions into plane-waves up to an
energy cut-off of 300 eV. We sample the Brillouin zone on an $8\times8\times1$
regular mesh, and if applicable, spin-orbit coupling (SOC) is
self-consistently included \cite{SOC_VASP}. Even though the stanene low-energy models are
extracted by projecting onto Sn $p_z$- and $sp_2$-type maximally
localized Wannier functions (MLWF) by using the WANNIER90 package
\cite{WANNIER90}, we compute the $\mathbb{Z}_2$ topological invariant following the
general method by Soluyanov and Vanderbilt \cite{WCC} as in the Z2Pack
code. That is, without relying on any pre-computed low-energy model, we directly compute the Wilson loop
of the entirely occupied ab-initio spectrum \cite{z2pack}.
Phonon dispersions are obtained within the context of the frozen phonon method \cite{Parlinski}
as implemented in the PHONOPY code \cite{PHONOPY} with a $2\times 2\times 1$ supercell (as explained in the next paragraph,
the unit cell consists already of $3\times 3$ SiC and $2\times 2$ stanene superstructures).

\begin{figure}[!t]
\centering
\includegraphics[width=\columnwidth,angle=0,clip=true]{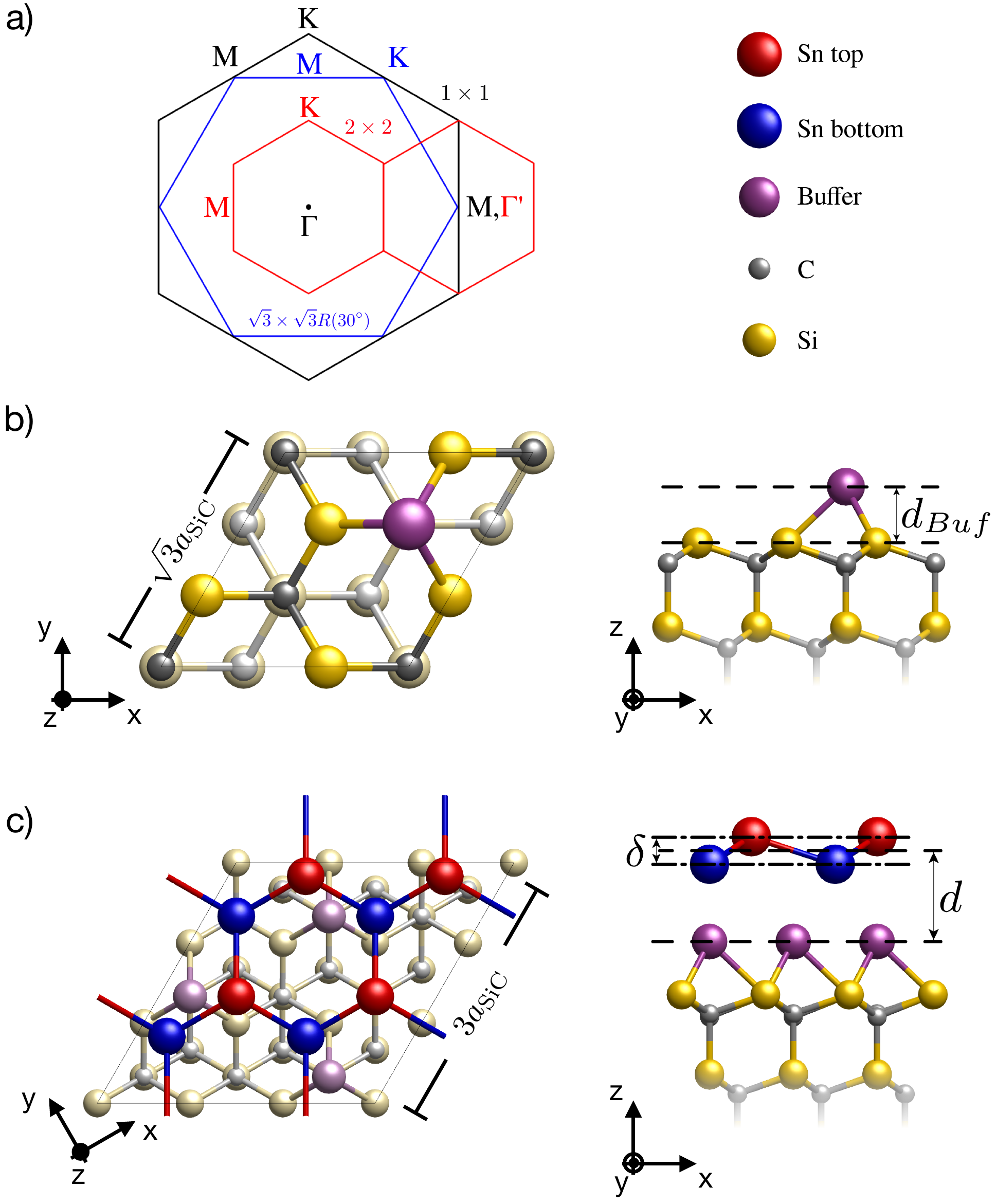}
\caption{a) Brillouin zones of $\sqrt{3}\times\sqrt{3} R(30^\circ)$ SiC
(blue), $1 \times 1$ stanene (black) and $2 \times 2$ stanene (red),
with relative high symmetry points. b) Top and side views of the
crystalline structure of buffered $\sqrt{3}\times\sqrt{3} R(30^\circ)$
SiC and c) $2\times2$ stanene/$3\times3$ buffered SiC. The bottom
termination of the slab models is artificially passivated by H atoms.
In the top view in panel c) the colors of the substrate are intentionally less intense compared
to panel b) in order to emphasize the hexagonal lattice of $2 \times 2$ stanene.}
\label{fig1}
\end{figure}

Each atom from group III and V can saturate three dangling bonds from
the SiC surface, either from Si atoms of the Si-terminated SiC or from C
atoms from the C-terminated SiC, as shown in Fig. \ref{fig1}b). This requires a
minimal $\sqrt{3}\times\sqrt{3} R(30^\circ)$ reconstruction, whose Brillouin zone and
crystal structure are reported in Fig. \ref{fig1}a).
Moreover, a $2\times2$ hexagonal lattice of stanene ($a=9.28$ \AA) turns out to have
a good commensuration with a $3\times3$ reconstruction of SiC
(lateral compressive strain $\varepsilon\sim0.62\%$), suggesting a promising
strain-free overlayer deposition. A $3\times3$ {\it buffered} SiC template ultimately
shows at its surface a triangular lattice of buffer atoms (see Fig. \ref{fig1}c).

\begin{figure}[!h]
\centering
\includegraphics[scale=0.4,angle=0,clip=true]{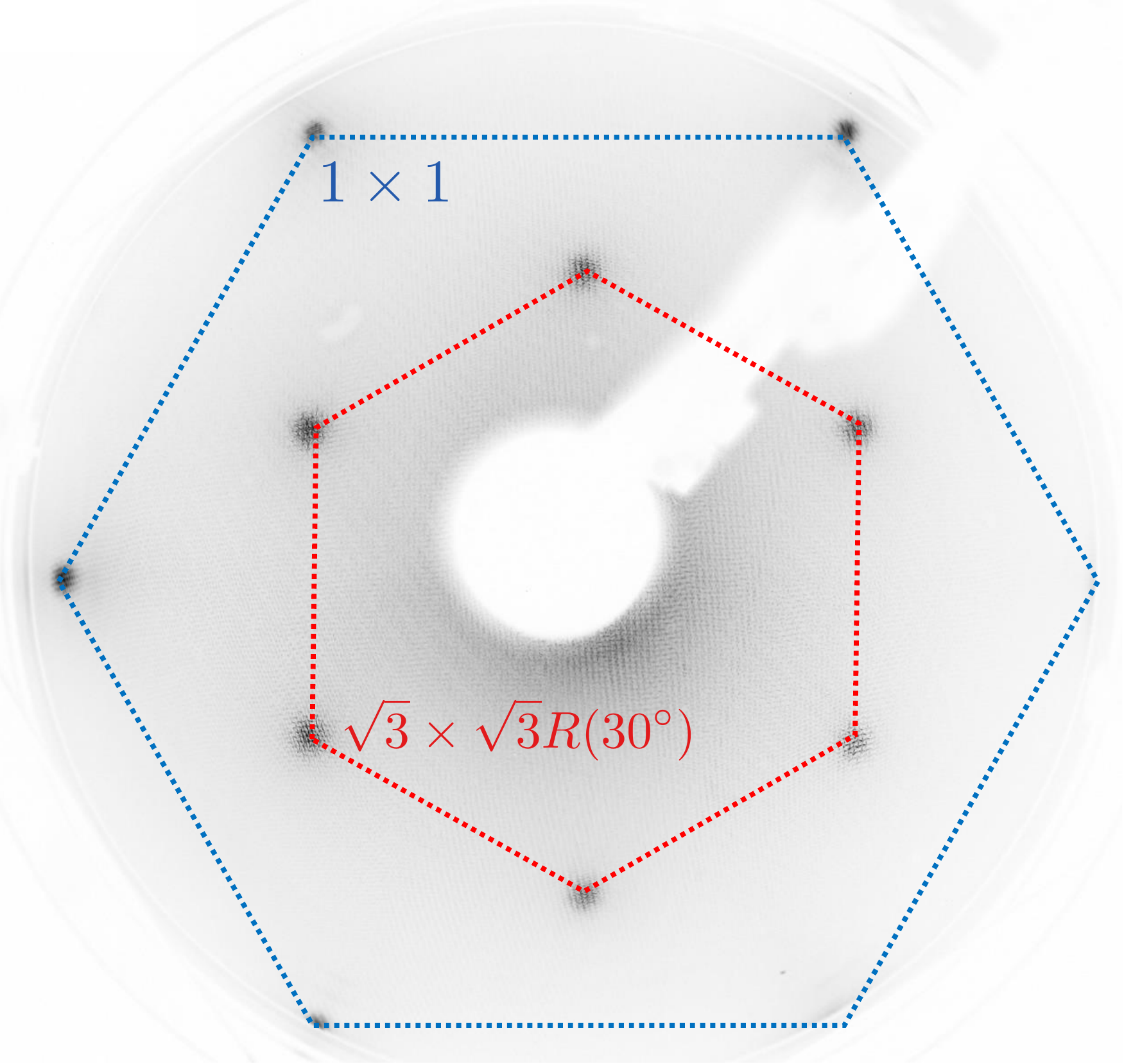}
\caption{LEED image of Al $\sqrt{3}\times\sqrt{3} R(30^\circ)$ on Si-terminated
4H-SiC(0001) recorded with an electron energy of 40 eV. Hexagons in red and blue
highlight the Al induced $\sqrt{3}\times\sqrt{3} R(30^\circ)$ and $1 \times 1$ spots, respectively.}
\label{figleeed}
\end{figure}

The large structural reconstructions enforce a folding of the electronic
states into the supercell Brillouin zones, which map onto the primitive
$1\times1$ Brillouin zones as sketched in Fig 1a. It is
usually simpler to achieve a transparent physical description in the
latter setting, where the unfolded band structure readily compares with
the freestanding models when the symmetry breaking induced by the
reconstruction is weak. The unfolding procedure we adopt in this work
follows the lines described in Refs.~\onlinecite{WeiKu,vaspunfold}.

For the experimental realization of a buffer layer we use $n$-doped (0.01-0.03 $\Omega$cm)
Si-terminated 4H-SiC(0001) wafer pieces. To prepare an atomically smooth, well-ordered
substrate surface on large-scales, the wafer pieces undergo a dry-etching process in a
helium diluted hydrogen atmosphere with a flow of 2 standard litre per minute at 950 mbar and temperatures
around 1200 $^\circ$C for roughly 10 minutes. Subsequently, the H-terminated SiC(0001) samples \cite{Glass2016}
were transferred to the preparation chamber (base pressure p $< 3\times 10^{-11}$ mbar) using a vacuum suitcase.
Surface quality was inspected in situ by low-energy electron diffraction (LEED).

Prior to the epitaxial growth of Al, the H termination of SiC(0001) has
to be removed. This is achieved by heating the substrate to $\sim 620
^\circ$C with a subsequent cooling to $\sim 350 ^\circ$C in the Al beam
of the effusion cell, to form the Al $\sqrt{3}\times\sqrt{3} R(30^\circ)$ lattice,
as shown in Fig. \ref{figleeed}, where LEED clearly shows the
reconstruction of the commensurate buffer layer. All samples were heated
by direct current and the temperatures were measured pyrometrically.


\section{Stanene on group-III buffer}
\begin{figure*}[!t]
\centering
\includegraphics[width=\textwidth,angle=0,clip=true]{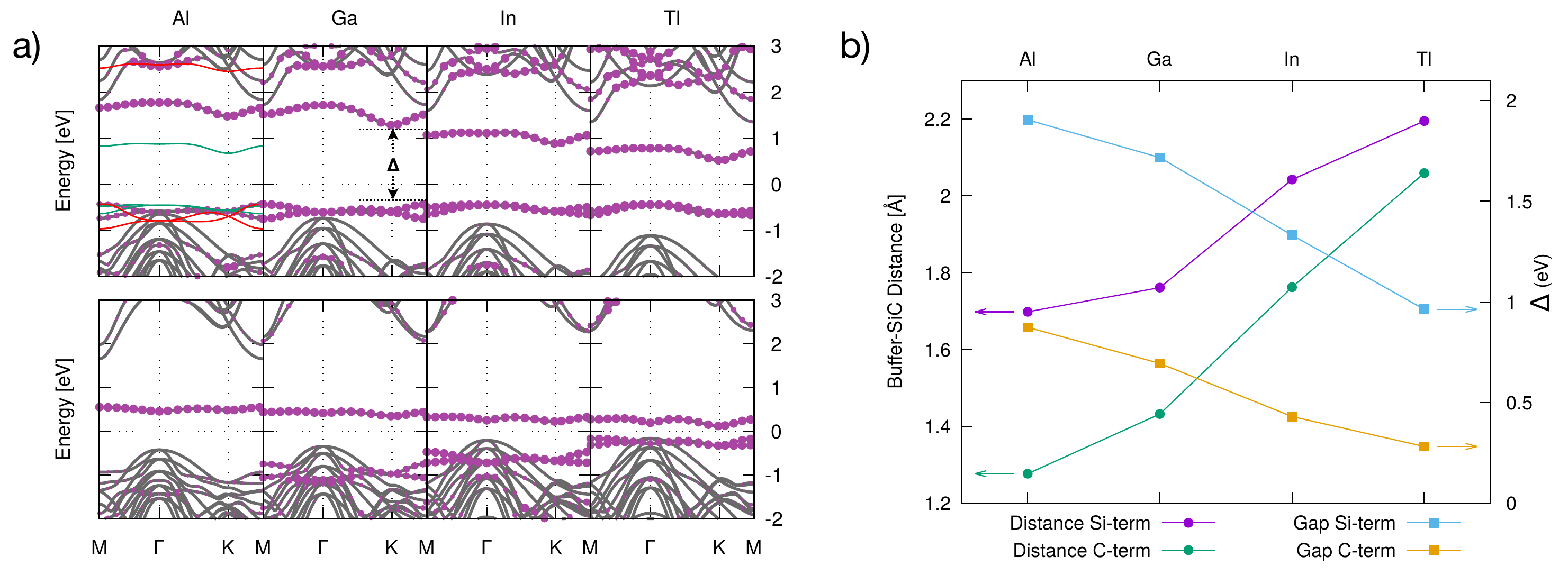}
\caption{a) Electronic band structures for group III buffered SiC (top row for Si-face and bottom row for C-face SiC) along the high-symmetry lines of the $\sqrt{3}\times\sqrt{3} R(30^\circ)$
Brillouin zone (see Fig. \ref{fig1}a). The coloured dots highlight the orbital contribution from the buffer atom.
For Al we additionally show the dependence of the bonding-antibonding in-gap states on the Al-SiC distance $d_\text{Buf}$: green corresponds to $d_\text{Buf}=2.1$\AA, i.e. roughly the same equilibrium distance obtained for Tl, while red corresponds to an exaggeratedly small value $d_\text{Buf}=1.3$\AA. 
b) Anticorrelation of the bonding-antibonding states gap with the distance between the group-III atom buffer layer and the SiC surface, shown for both substrate terminations (Si- and C-faces).}
\label{fig2}
\end{figure*}
\subsection{Buffer analysis}\label{sec:bufIII}
We consider Al, Ga, In, and Tl as group III elements of the periodic
table, which all share the same $s^2p^1$ valence electronic configuration. 
It has been shown that Al and other group III atoms can saturate the (111) surface of a silicon crystal,
inducing a $\sqrt{3}\times\sqrt{3} R(30^\circ)$ reconstruction \cite{Lander,Northrup,Nicholls,Hamers}. 
The underlying mechanism is that the three Al valence orbitals saturate
three of the Si surface dangling bonds, which naturally leads to a
$\sqrt{3}\times\sqrt{3}$ coverage of the Si substrate.
By analogy, and based on the similarities that the SiC electronic structure
shares with Si, we assume here that the deposition of group-III atoms
induces on the SiC surface the same $\sqrt{3}\times\sqrt{3} R(30^\circ)$
reconstruction as on silicon. Our theoretical assumption is indeed experimentally proven here
for the first time for the case of an Al buffer on SiC. Fig. \ref{figleeed} shows
a clear $\sqrt{3}\times\sqrt{3} R(30^\circ)$ reconstruction (red hexagon) on
an underlying $1\times 1$ template (blue hexagon). 

The $sp^3$ orbitals of Al on SiC host the three valence electrons. 
The fourth one is empty and of $sp_z$-type character and gives rise to an antibonding in-gap state.
The resulting band structures for this type of buffer layers on SiC (Si-face) are shown in Fig. \ref{fig2}a). The energy position
of the antibonding in-gap state anticorrelates with the bonding distance
of the buffer atom to SiC, as we show in Fig. \ref{fig2}b). The lighter the buffer atom
the shorter is the resulting distance, with the concomitant upward level
repulsion of the antibonding orbital. This anticorrelation trend is
independent of the SiC termination, holding both for the Si- and for the C-terminated
surface.

In Fig.~\ref{fig2}a, in the case of the Al buffer on the Si face, in addition to the equilibrium Al-SiC distance $d_\text{Buf}\!=\!1.7$\AA, we show the bonding-antibonding in-gap states with predominant buffer character for two more values of $d_\text{Buf}$. 
If we lift Al to distances close to those obtained for Tl, the bonding-antibonding states are located at a very similar position with a comparable gap $\Delta$ (green curve). On the other hand, by pushing the buffer artificially close to SiC ($d_\text{Buf}\!=\!1.3$\AA, red curve), $\Delta$ increases, with the antibonding state merging into the SiC conduction band. This evidences a rather weak dependence of the buffer states on the atomic species, proving that they instead decisively depend on $d_\text{Buf}$. 

The energy position of the antibonding buffer level, and in turn the excitation gap,  affect the strength of the hybridization when we add the stanene monolayer on top:
The larger the gap of the SiC+buffer system is, the more strongly the buffer (conduction and valence) states hybridize with SiC.
This implies that for a large $\Delta$, the corresponding buffer states penetrate more pronouncedly into the bulk, and their surface localization is reduced. Stanene will hence be less affected in this situation.
In the opposite case of a narrow gap, with buffer states situated at low energies, these have more weight at the surface and we expect sizable interaction effects between the buffer and stanene.
A further important consequence is that top and bottom Sn atoms of the buckled inversion-breaking geometry have significantly distinct energy levels, as they are differently exposed to a strong interface potential.
In Sec.~\ref{sec:semenoff} we are going to show explicitly that this directly translates into the presence of a staggered potential term (Semenoff mass). This is known to be detrimental to the QSH phase, as it tends to compete with the second-nearest neighbor SOC \cite{KM1}.

\begin{table}[!b]
\centering
\caption{Summary of the DFT results obtained for buffer layers made of group III elements (for a geometry of stanene with preserved hexagonal symmetry).
Si and C refer to the silicon and carbon terminated SiC, respectively. $\Delta {\text E}_{\text K}$ is the
stanene energy gap at the K point. The dash symbol indicates that the system is metallic. $d_\text{Buf}$, $d$ and $\delta$
are the buffer layer-SiC distance, the stanene-buffer distance and the buckling height, respectively. 
$\lambda_v$ and $3\sqrt{3}\lambda_\text{SO}$ are the Semenoff mass and the effective SOC
entering the Hamiltonian at the K point as given in Eq. \ref{eq3}, respectively.
We also report the values of the $\mathbb{Z}_2$-invariant, even though in the case of group III this is always 0 at $d \! = \! d_\text{eq}$.
The calculations are performed within GGA-PBE \cite{PBE}.
In the trivial $\mathbb{Z}_2 \! = \! 0$ phase, $\Delta {\text E}_{\text K} = 2[\lambda_v - 3\sqrt{3}\lambda_\text{SO}]$.}
\label{tab1}
\begin{tabular}{p{1.0cm}p{2.5cm}p{1cm}p{1cm}p{1cm}p{1cm}}
\\
\hline\hline
        &   &  Al  &  Ga  &  In  &  Tl \\
\hline        
Si      &   &      &      &      &     \\
        & $\Delta {\text E}_{\text K} (\text{meV})$ &  56    &  66    &  60    &  63   \\
        & $d_\text{Buf} (\text{\AA})$                    &  1.7   &  1.8   &  2.0   &  2.2  \\
        & $d (\text{\AA})$                          &  2.9   &  2.9   &  3.0   &  3.1  \\
        & $\delta (\text{\AA})$                     &  0.44  &  0.48  &  0.48  &  0.48 \\
        & $\mathbb{Z}_2$                            &  0     &  0     &  0     &  0    \\
        & $\lambda_v (\text{meV})$                  &  49    &  58    &  50    &  52   \\
        & $3\sqrt{3}\lambda_\text{SO} (\text{meV})$ &  21    &  25    &  20    &  21   \\
\hline
C       &   &      &      &      &     \\
        & $\Delta {\text E}_{\text K} (\text{meV})$ &  -     &  -     &  -     &  -    \\
        & $d_\text{Buf} (\text{\AA})$                     &  1.3   &  1.4   &  1.8   &  2.1  \\
        & $d (\text{\AA})$                          &  2.8   &  2.8   &  3.0   &  3.1  \\
        & $\delta (\text{\AA})$                     &  0.44  &  0.48  &  0.48  &  0.48 \\
        & $\mathbb{Z}_2$                                     &  -     &  -     &  -     &  -    \\        
\hline
\end{tabular}
\end{table}

\subsection{Group-III buffered SiC + stanene}\label{sec:stanIII}
\begin{figure*}[!t]
\centering
\includegraphics[width=\textwidth,angle=0,clip=true]{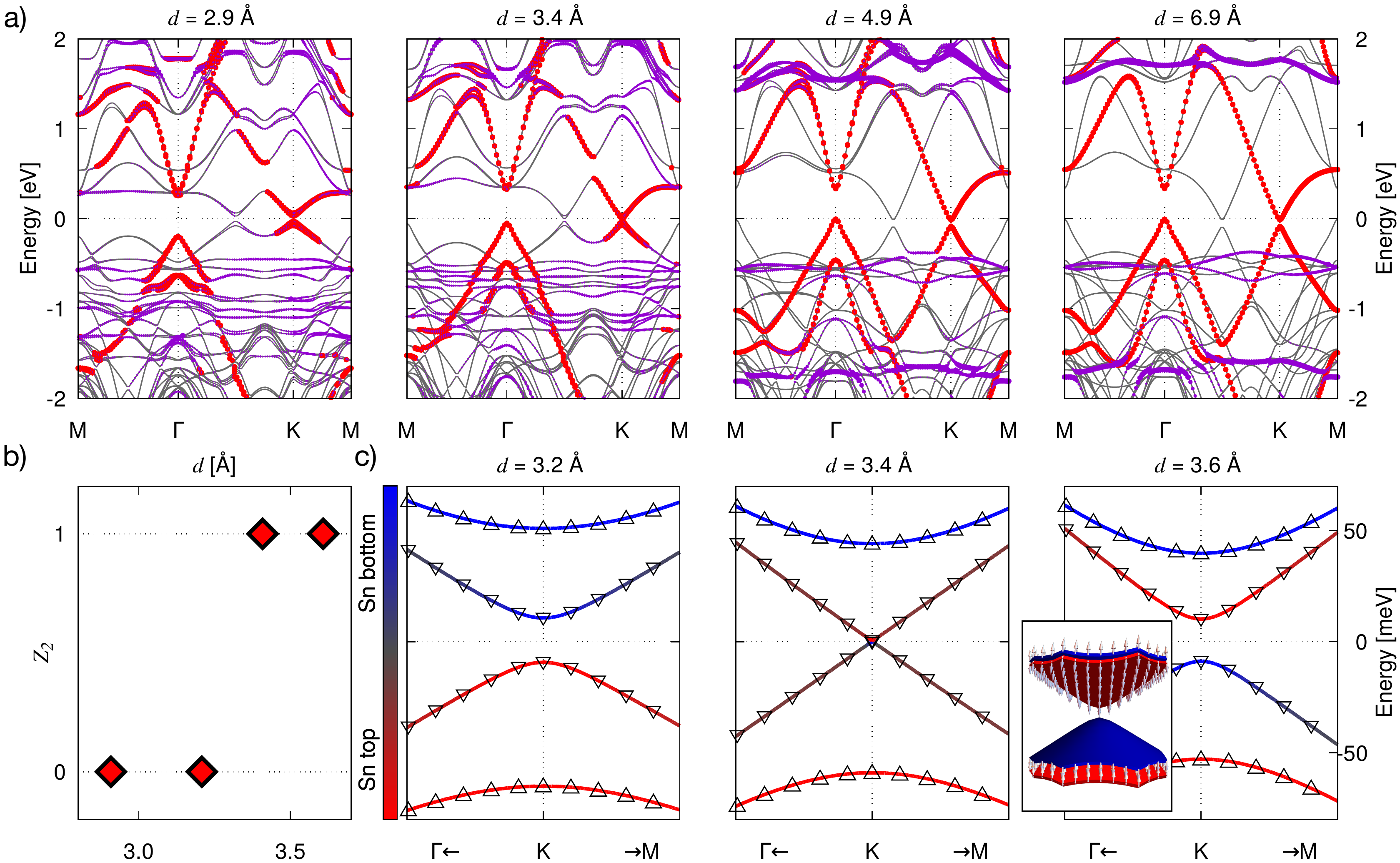}
\caption{a) Evolution of the stanene/Al/SiC electronic band structure at different
stanene-Al buffer distances ($d = 2.9$\AA, $3.4$\AA, $4.9$\AA\, and $6.9$\AA\, from left to right panels).
Red dots highlight the unfolding weight from the $2\times2$ Brillouin zone onto the primitive $1\times1$,
while the violet color code refers to the orbital contribution from the buffer atoms.
b) Behaviour of the $\mathbb{Z}_2$ topological invariant close to the critical transition distance $d_\text{cr}$.
c) Zooms around the K point across the topological transition. Red and blue colors
refer to orbital character from top and bottom Sn atoms, respectively.
In the inset we report a three-dimensional view of the spin-texture. The in-plane $S_x$ and $S_y$
components are negligible compared to the out-of-plane $S_z$ component.}
\label{fig3}
\end{figure*}

We now analyze the behavior of stanene, in particular its geometrical and its topological properties, when deposited on the group-III buffer layer. For the sake of clarity, we derive most of the following conclusion by focusing on the specific case of the Al-buffer on SiC.
The main purpose is to understand the physical mechanisms behind the interaction between stanene and the buffered substrate. 
With this in mind, we monitor the evolution of the electronic band
structure of stanene as its bonding distance is artificially tuned from
an unrealistically large value down to the equilibrium one
$d_\text{eq}\sim2.9$\AA\, (see Fig. \ref{fig3}a). The $2\times2$
structure reconstruction induced by the lattice commensuration leads to
band foldings \cite{WeiKu}. When the stanene-substrate distance is large
enough to reduce the hybridization effects and the Coulomb interaction, we recover the
freestanding stanene electronic properties. By unfolding the stanene
band structure from the $2\times2$ Brillouin zone into the primitive $1\times1$, we
can directly compare with the results of the freestanding description
\cite{Stanene}. The reduction of the distance towards the equilibrium geometry
and the concomitant interaction causes the opening of hybridization
gaps in the stanene band structure. Along with it, we find a distribution (spread) of stanene
character over the Al/SiC electronic states. When the distance becomes
smaller than the critical value $d_\text{cr}\sim3.2$\AA, we observe a switching of the $\mathbb{Z}_2$
invariant from 1 to 0 (Fig. \ref{fig3}b). This evidence marks the
transition from freestanding-like topologically non-trivial to
topologically trivial stanene, at least within GGA framework.
As we will see in the following, the same
mechanism leads to a different outcome in the case of group V.

A change in the topological
invariant can only occur through an inversion of the bulk gap \cite{RPMHasan,RPMQi}. This evolution is
indeed what we observe and show in Fig. \ref{fig3}c. First of all, 
the breaking of the inversion symmetry induced by the presence of
the substrate removes the band degeneracies at the K points. This
effect goes under the name of valley-contrasting physics, or
valleytronics, and is a well-known property of gated graphene and
transition metal dichalcogenides \cite{GrapheneValley,MoS2Valley}. The gap closure is accompanied
by a change in the sublattice character of the bands. In fact, in the
topological (trivial) phase, the contribution from the bottom Sn (top
Sn) atom dominates the valence band maximum, and vice-versa for the
conduction band minimum. The closure of the gap occurs through a linear
Dirac-like band touching where the sublattice character is equally
mixed, as highlighted by the brownish color of the linear branches in the middle panel of Fig. \ref{fig3}c). 

In the next section, we will establish an analogy with the
topological transition as described by the Kane and Mele model \cite{KM1}, and
analyze the quantitative role of the Semenoff mass. Note that the trend observed here for stanene on
Al/SiC is common to all the group-III buffer setups, regardless of the SiC
termination (see Table \ref{tab1}). 
As shown in Fig. \ref{fig2}b, it holds that the heavier the buffer atom deposited on SiC, the smaller the bulk gap. For this reason, the most favorable group-III buffer for growing quasi-freestanding stanene is Al. The latter leads, however, to a topologically trivial configuration for stanene, demonstrating how going downwards in the ``triels'' group of the periodic table is not a promising approach to obtain $\mathbb{Z}_2\!=\!1$. 


\section{Role of the Semenoff mass}\label{sec:semenoff}
Interpreted within the framework of the Kane and Mele model \cite{KM1}, our tight-binding Hamiltonian reads 
\begin{eqnarray}
\label{eq1}
H = &&t\sum_{\langle ij\rangle,\alpha} c_{i,\alpha}^\dag c_{j,\alpha} + i\lambda_\text{SO}\sum_{\langle\langle ij\rangle\rangle,\alpha\alpha'}\nu_{ij}c_{i,\alpha}^\dag s^z_{\alpha\alpha'}c_{j,\alpha'} \nonumber \\
&& + i\lambda_{R}\sum_{\langle ij\rangle,\alpha\alpha'}c_{i,\alpha}^\dag ({\mathbf s}\times \hat{{\mathbf d}}_{ij})^z_{\alpha\alpha'}c_{j,\alpha'} \nonumber \\
&& + \lambda_v\sum_{i,\alpha} \xi_i c_{i,\alpha}^\dag c_{i,\alpha}
\end{eqnarray}
where the first is a nearest neighbor hopping term on the honeycomb lattice and the second 
a mirror symmetric SOC one (here $\nu_{ij}\hat{{\mathbf u}}_z = (2/\sqrt{3})(\hat{{\mathbf d}}_1 \times \hat{{\mathbf d}}_2)$ where $\hat{{\mathbf d}}_{1,2}$ are unit vectors along
the two bonds from site $j$ to site $i$, and $\hat{{\mathbf u}}_z$ is a unit vector perpendicular to the stanene plane), and ${\mathbf s}$ are the Pauli matrices for the electron spin. 
The third term is a nearest-neighbor Rashba coupling due to a perpendicular electric field
or to an interaction with a substrate and the last term sets the staggered potential (Semenoff
mass) $\lambda_v$ ($\xi_i=\pm 1$ depending on the sublattice). The latter differentiates the on-site energies of the two atoms
constituting the bipartite honeycomb lattice. Without the Rashba term, the momentum-space Hamiltonian assumes the form
\begin{eqnarray}
\label{eq2}
H({\mathbf k}) & = & t(1+2\cos{x}\cos{y})\Gamma_1 -2t\cos{x}\sin{y}\Gamma^{12} \nonumber \\
& + & \lambda_\text{SO}(2\sin{2x} - 4\sin{x}\cos{y})\Gamma^{15} + \lambda_v\Gamma_2
\end{eqnarray}
with $x=k_xa/2$ and $y=\sqrt{3}k_ya/2$ ($a$ being the hexagonal unit cell lattice constant, i.e. $\overline{\langle ij \rangle} = a/\sqrt{3}$),
$\Gamma_1=\sigma_x\otimes s_0$, $\Gamma_2=\sigma_z\otimes s_0$ and $\Gamma_5=\sigma_y\otimes s_z$ three of the five Dirac matrices
with $\sigma_i$ describing the sublattice degree of freedom and $\Gamma^{ab} = [\Gamma_a,\Gamma_b]/2i$.
At the K point, the Hamiltonian is diagonal in the sublattice basis ${A\!\uparrow, B\!\uparrow, A\!\downarrow, B\!\downarrow}$:
\begin{eqnarray}
\label{eq3}
\begin{bsmallmatrix}
    \lambda_v - 3\sqrt{3}\lambda_\text{SO} & 0 & 0 & 0 \\
    0 & -(\lambda_v - 3\sqrt{3}\lambda_\text{SO}) & 0 & 0 \\
    0 & 0 & \lambda_v + 3\sqrt{3}\lambda_\text{SO} & 0 \\
    0 & 0 & 0 & -(\lambda_v + 3\sqrt{3}\lambda_\text{SO})
\end{bsmallmatrix}
\end{eqnarray}
The two $\uparrow$- and the two $\downarrow$-eigenvalues with smaller
and larger splitting, respectively, give perfect account of the four
bands visible in the actual DFT calculation (Fig. \ref{fig3}c). The
matching between the DFT result and the Kane-Mele model can be pushed
further upon comparing the sublattice character of these four
eigenvalues and its evolution with $d$. We define $d_\text{cr}$ as the
distance at which the two lowest-lying DFT eigenvalues touch each other.
For $d<d_\text{cr}$ the character of both the uppermost eigenvalues is
on one sublattice, namely the one corresponding to the bottom Sn-atom,
while both eigenvalues in valence band belong to the top Sn. For
$d>d_\text{cr}$ the sublattice character at the K point is instead
interchanged, a situation which in the mapping onto the Kane-Mele is obtained in the QSH
phase for values of the Semenoff mass $\lambda_v <
3\sqrt{3}\lambda_\text{SO}$\cite{KM1}.
In the topological phase with inverted sublattice character, the lattice
site with the highest on-site energy ($+\lambda_v$) contributes to
the valence band maximum at K, at odds with the trivial atomic
limit. 

\begin{figure*}[t!]
\centering
\includegraphics[width=\textwidth,angle=0,clip=true]{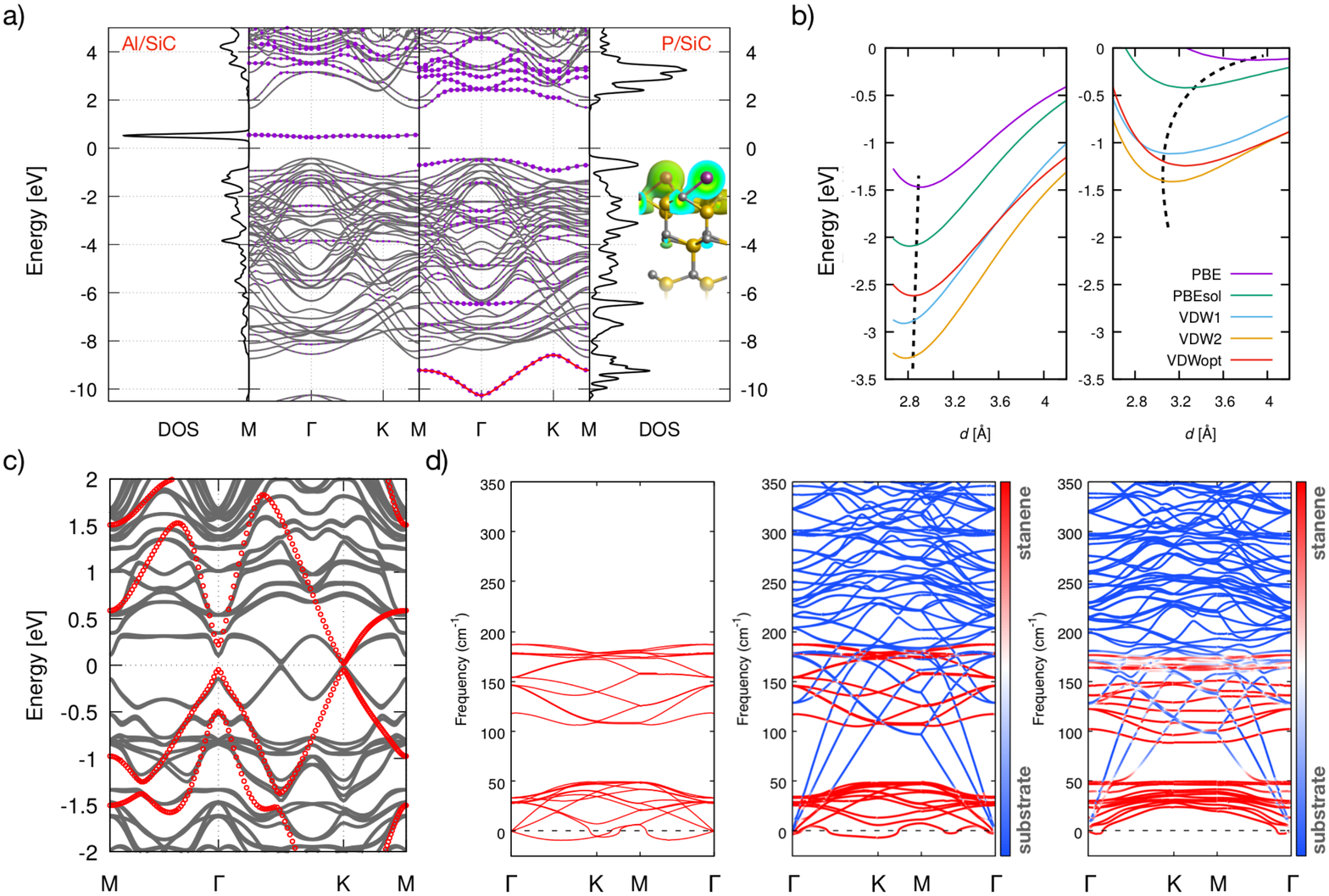}
\caption{a) Buffer resolved density of the states and band structure of Al/SiC and P/SiC (C-face hereafter in this caption), respectively.
The dots in the band structure highlight the orbital weight of buffer atoms. In the inset, the charge distribution projected on the
states forming the lone pair around 9 eV of binding energy for P/SiC is shown.
b) Binding energy curves for stanene on Al Si-terminated SiC (left panel) and on P C-terminated SiC (right panel) with and without different flavours of van der Waals
corrected functionals (PBE \cite{PBE}, PBEsol \cite{PBEsol}, VDW1 \cite{VDW1}, VDW2 \cite{VDW2}, VDWopt \cite{VDWopt}). The black dashed lines are guide to the eye to track the evolution of the equilibrium distance within the
tested approximations.
c) band structure (with VDW1 van der Waals correction) of $2 \times 2$ stanene on P/SiC with the red dots referring to freestanding stanene in the primitive $1 \times 1$
structure. d) Phonon spectra of freestanding stanene (leftmost panel) and stanene on P/SiC at the $3.9\text{\AA}$ PBE (central panel) and $2.9\text{\AA}$ VDW1 (rightmost panel)
equilibrium distances. The color code refers to pure P/SiC (blue) and pure stanene (red) contributions to the phonon modes.}
\label{fig4}
\end{figure*}

Let us note that the inversion of sublattice character is well defined
only at the K and K$'$ points, where the Kane-Mele Hamiltonian is
diagonal in the original basis. A suitable definition of the character
over the whole Brillouin zone can be obtained by rotating the Kane-Mele Hamiltonian
at each $k$-point in the basis of the eigenvectors of~(\ref{eq2})
calculated at $\lambda_\text{SO}\!=\!0$. At the $\Gamma$ point, for
instance, the Kane-Mele model is made of two identical 2$\times$2
blocks, with $\pm \lambda_v$ on the diagonal and $3t$ between the $A$ and
$B$ sites. Going away from the $\Gamma$ point and continuing to use the
``bonding''-``antibonding'' basis defined above, the SOC acquires
off-diagonal elements which mix the ``bonding''-``antibonding''
eigenvectors approaching K and K$'$ leading to the inversion at the
topological phase transition.

So far we have discussed the qualitative correspondence between the
Kane-Mele model and the DFT results. Now, we make a more quantitative
comparison by estimating the Semenoff mass and SOC term via a fit
of the DFT eigenvalues at the K point with the analytic expressions given in Eq. \ref{eq3}. 
For the Al buffer, at $d_\text{eq}$, we extract a difference between the $p_z$ onsite energies of the top- and
bottom-Sn atoms of $2\lambda_v = 0.098$ eV. The highest levels belong to
the bottom Sn-atoms, that feel the buffer layer stronger than the top
ones and are hence pushed up in energy. In the Kane-Mele Hamiltonian $\lambda_\text{SO}$
represents a non-local term which accounts for the coupling of orbitals
with parallel spin on the same sublattice. However, a finite buckling
angle introduces a non-vanishing local contribution to the SOC \cite{Silicine}. We estimate
that, in stanene on Al/SiC, $3\sqrt{3}\lambda_\text{SO}\sim 21$ meV. 
Therefore, the relationship $\lambda_v > 3\sqrt{3}\lambda_\text{SO}$
holds for stanene on the Al buffer at the equilibrium distance,
confirming the trivial ground state found in our DFT computations.
As can be inferred from Table \ref{tab1}, the same relationship holds
for all the investigated group-III buffers.

Let us briefly comment on the quantitative estimate of the Semenoff
mass: an alternative way of extracting the staggered potential is from the difference of the
$p_z$-local levels of the two sublattices via a Wannier projection of the DFT band-structure.
In order to get a satisfactory agreement with DFT, the projection must contain all
the $p$-orbitals of tin as well as orbitals from the buffer atoms. This
means that the resulting Wannier Hamiltonian is defined on a much larger Hilbert
space than the one of the Kane-Mele model, which assumes one single
orbital per site. To directly connect to $\lambda_v$, one would
have to further downfold the Wannier Hamiltonian onto a $p_z$-type
low-energy model. A direct projection onto such a minimal model, however,
turns out to yield a poor correspondence with the DFT bands.
Furthermore, there is another argument in favor of an extended local
basis, namely that this allows to disentangle hybridization
effects stemming from the buffer. As described above, we adopt the simpler procedure based on the 
fit of the DFT eigenvalues at the K point to the analytic results of Eq.~\ref{eq3}.

In the previous analysis, we have neglected the Rashba coupling in Eq. \ref{eq1}.
The presence of a finite $\lambda_{R}$ mixes the $\uparrow$- and $\downarrow$-blocks
of the Hamiltonian breaking the conservation of the $z$-component of the spin. 
Moreover, a chiral spin-texture appears around the K point.
We checked that this effect is negligible, as shown in the inset to Fig.~\ref{fig3}c,
proving that our assumption $\lambda_{R} \! =\! 0$ is justified and does not affect
the qualitative arguments given above. Let us additionally note that, as recently demonstrated,
the impact of the Rashba coupling in 2D superstructures can be reduced by a proper
choice of the lattice commensuration \cite{Roche}.

\section{Stanene on group-V buffer}

\begin{table}[!b]
\centering
\caption{Same as Table \ref{tab1} but for buffer layers made of group-V elements.
The column on the left refers to Si-face or C-face of the SiC substrate.
The asterisk (*) in the entries of $\delta$ indicates that the stanene low-buckled hexagonal geometry
is not preserved anymore and it is not possible to define a buckling parameter.
The calculations are performed within the VDW1 framework \cite{VDW1}.
In the topological $\mathbb{Z}_2\!=\!1$ phase, $\Delta {\text E}_{\text K}\! =\! 2[3\sqrt{3}\lambda_\text{SO} - \lambda_v]$.}
\label{tab2}
\begin{tabular}{p{1.0cm}p{2.5cm}p{1cm}p{1cm}p{1cm}p{1cm}}
\\
\hline\hline
        &   &   P  &  As  &  Sb  &  Bi \\
\hline        
Si      &   &      &      &      &     \\
        & $\Delta {\text E}_{\text K} (\text{meV})$ &  -     &  -     &  -     &  -    \\
        & $d_\text{Buf} (\text{\AA})$                    &  1.8   &  2.0   &  2.1   &  2.2  \\
        & $d (\text{\AA})$                          &  1.9   &  1.7   &  3.0   &  3.2  \\
        & $\delta (\text{\AA})$                     &  *     &  *     &  0.40  &  0.43 \\
        & $\mathbb{Z}_2$                            &  -     &  -     &  -     &  -    \\        
\hline
C       &   &      &      &      &     \\
        & $\Delta {\text E}_{\text K} (\text{meV})$ &  13    &  34    &  1     &  13   \\
        & $d_\text{Buf} (\text{\AA})$                    &  1.2   &  1.4   &  1.6   &  1.8  \\
        & $d (\text{\AA})$                          &  2.9   &  3.0   &  3.2   &  3.2  \\
        & $\delta (\text{\AA})$                     &  0.43  &  0.42  &  0.42  &  0.42 \\
        & $\mathbb{Z}_2$                            &  1     &  1     &  1     &  1    \\
        & $\lambda_v (\text{meV})$                  &  16    &  6     &  19    &  10   \\
        & $3\sqrt{3}\lambda_\text{SO} (\text{meV})$ &  23    &  23    &  19    &  17   \\
\hline
\end{tabular}
\end{table}

Equipped with the microscopic understanding gained in the previous
section of the mechanism responsible for the topological transition in
stanene on the buffer layer, we shall now address the key question of our
study: how to realize quasi freestanding topologically non-trivial stanene. 
From our analysis of group III, we have learned that the buffer in-gap states have to 
stay as far as possible away from $E_F$ and, at the same time, the Semenoff mass term must be minimized. 
To limit the detrimental effect of these antibonding states, the corresponding bonds ought to be saturated.
Two more valence electrons in buffer exactly serve this purpose, offered by
the $p$-shell of group-V atoms.

In Fig. \ref{fig4}a we compare the density of the states (DOS) and
band structure of Al/SiC and P/SiC (C-face). From the comparison between
group-III and V buffer it is clear that our expectation of the quenching
of the antibonding state is fulfilled: for group V we indeed observe a
chemically inert lone pair characterized by a pronounced $s$-type
orbital character (see inset in Fig. \ref{fig4}a). This is located around $10$ eV
below the Fermi level, leaving as a consequence more ``room'' within the
gap for an interference-free positioning of the stanene layer. 
Thanks to the reduced monolayer/substrate interaction, the induced
Semenoff mass is indeed smaller. This is illustrated in Table
\ref{tab2}, where we summarize our results for stanene on group-V
buffer/SiC substrates. 
All the calculated values of $\lambda_v$ are sensibly reduced compared to the ones 
reported in Table \ref{tab1}. 
Furthermore, contrary to group III, in the case of group V buffers the hexagonal symmetry of stanene is 
preserved during the relaxation, reflecting the much weaker hybridization strength.

Before moving to the analysis of the resulting electronic properties, we turn our
attention to the nature of the bonding between stanene and the buffer layer.
To this purpose, in Fig.~\ref{fig4}b we show the binding energies of stanene on
Al (left panel) and P (right panel) buffer/SiC within different levels of sophistication,
explicitly including long-range dispersion to account for 
van der Waals interactions. The difference in the two behaviour suggests a distinct
nature of the bonding. While on group III buffer/SiC van der Waals corrections
do not change qualitatively the equilibrium geometry of stanene, a sizeable reduction of the
equilibrium distance occurs for group V buffer/SiC. Such an evidence hints at a strong covalent nature of the bonding on group-III
buffer, as a result of the hybridization between stanene
and the antibonding in-gap state. On the other hand, the inert lone pair prevents from any form of
chemical bonding, leaving room for the predominance of the van der Waals interaction.   

The band structure of stanene on C-terminated P/SiC at the van der Waals equilibrium geometry is shown in Fig.~\ref{fig4}c.
The Dirac physics around the K point resembles to a large extent
the one of freestanding stanene \cite{Stanene}. Our conclusion is further
substantiated by the explicit computation of the $\mathbb{Z}_2$ invariant (not shown),
where a Kramers' pair switching indicates the non-trivial topological nature
($\mathbb{Z}_2 = 1$) \cite{WCC}.

Our calculations reveal that P and As buffer-atoms on C-terminated
SiC are ideal templates to host stanene in the QSH phase. 
The high electronegativity of carbon plays an important role. The resulting strengthening of the bonding of the buffer atoms to the substrate,
and in turn of the hybridization between the in-gap bands and the continuum of SiC electronic states, sets favourable
conditions for a topological ground state. 
We predict stanene to be a QSH on Bi as well, but it is likely that Bi on SiC
prefers the 2/3 coverage, as recently reported in the case of bismuthene/SiC \cite{Reis287}. 
Stanene on Sb/SiC is right at the verge of the topological transition, being the gap at the K point not larger than $\sim 1$ meV. 
The configurations on the other termination of SiC are instead metallic and in the cases of P and As, also highly distorted. 
Being less electronegative than C, Si
attracts less negative charge from the buffer atoms compared to what C
does, resulting in turn in a larger SiC/buffer distance $d_\text{Buf}$ (see Table \ref{tab2}).
This confirms that the C-face, in combination with group-V buffer layers, offers the ideal degree of hybridization to host stanene without spoiling the QSH phase.

Having assessed the possibility to achieve a topological ground state for stanene,
an important issue is yet to be addressed. It is known that freestanding stanene as predicted
in Ref.~\onlinecite{Stanene} is not dynamically stable in the low-buckled (LB) geometry of interest
here \cite{Tang}. This instability is a consequence of negative phonon frequencies
that the flexural ZA mode has along the $\Gamma$-K and $\Gamma$-M directions, as shown
in the leftmost panel of Fig. \ref{fig4}d). The origin of negative frequencies 
is ascribed to the weak $\pi$-$\pi$ bonding of LB stanene that fails to stabilize a
buckled configuration. It is interesting to note how a dumbbell (DB) stanene configurations
has been theoretically proposed to be stable and topologically non-trivial on a variety of different
substrates \cite{Tang}. At odds with LB stanene, however, in DB stanene the band inversion induced
by the SOC occurs at the $\Gamma$ point, and the Dirac-like physics at the K point is irredeemably spoiled.

The substrate engineering based on our buffer strategy turns out to be decisive to stabilize
stanene in the LB configuration. As shown in the central and rightmost panels of Fig. \ref{fig4}d),
the interaction with the substrate reduces the negative phonon frequencies,
eventually ending up in a stable configuration at the van der Waals equilibrium distance.
In such a configuration, all the frequencies of phonon modes associated to stanene are positive.    
The small still imaginary frequencies in the long-wavelength ${\bf q}\rightarrow 0$ flexural acoustic ZA mode is a
well known artifact that is not related to a structural instability \cite{LiuZA,Peng,Yu}.
The stabilization of the ZA phonon mode via interaction with the buffer is a result of the
very nature of this mode, which in atomically thin systems involves a vertical bending of the layer, i.e.
a flexural mode in the direction of the substrate. We cannot exclude at this point a high-buckled (HB)
configuration of stanene, since it has been reported that for freestanding tin and lead monolayers, a metallic HB phase
is energetically favourable against the LB one \cite{Rivero}. However, we stress again here that 
the role of the substrate, within the concept we propose in this work, is pivotal to accomplish
a quasi-freestanding LB phase, which is dynamically stable but still retains a QSH physics at the K point.    

\section{Conclusions}

Through a systematic first-principles analysis we have unveiled the
mechanisms underlying the delicate QSH formation in stanene 
grown on a SiC(0001) substrate, and how these growth conditions can be
fundamentally improved by an appropriate buffer between substrate and monolayer.
We followed the guiding principle that a technologically
relevant substrate is crucial for any future integration of QSH
systems in novel spin-based devices. SiC distinguishes itself as
one of the most promising substrate candidates, we have accomplished a
substrate engineering mechanism that employs a buffer layer for the
saturation of the substrate dangling bonds. We have further revealed the microscopic processes
that are deleterious to the formation of the topological phase, owing to the electronic hybridization between
the substrate and the stanene monolayer. Combining these insights,
we predict a stable QSH phase of stanene in the four cases of group-V buffered SiC.

By analyzing the hybridization between the stanene layer and the buffer,
and by mapping the full ab-initio Hamiltonian onto a generalized effective
Hamitonian including a staggered potential (Semenoff mass term), we show that some buffer layers are more suitable than others to
protect the QSH formation in the monolayer. The strategy we follow then is to minimize the detrimental staggered
potential by choosing such atoms which have their bonding and
antibonding states energetically far away from the chemical potential.
This leads us to concrete suggestions for promising buffer materials,
in particular the use of group-V elements (P and As).

Note that our theoretical study only establishes a lower bound to the
potential range of parameter space in which the QSH phase in stanene might be stabilized by proper substrate engineering.
This is because our DFT approach utilizes the GGA approximation,
which, due to the prohibitive
size of the reconstructions we considered,
would be extremely time-consuming to replace it by alternative
procedures. The use of GGA suggests that a $\sim 20\%$ underestimation
of the antibonding states energy position is possible (see Fig. \ref{fig2}a), and that some
of the trivial configurations we have found (Table \ref{tab1} and
\ref{tab2}) are in fact still in the QSH domain.     
In terms of an explicit analysis of the buffer-assisted growth of
stanene on SiC, we reported the first detailed LEED pattern
(Fig. \ref{figleeed}) demonstrating a $\sqrt{3}\times\sqrt{3} R(30^\circ)$ reconstruction of SiC after Al deposition.
Together, our efforts constitute promising steps towards the
accomplishment for the realization of quantum spin Hall candidate
materials through buffer engineering.

\section*{Acknowledgement}
The authors acknowledge G. Profeta for fruitful and inspiring discussions.
This work was supported by the DFG through SFB1170 "ToCoTronics" and by ERC-StG-336012-Thomale-TOPOLECTRICS.
We gratefully acknowledge the Gauss Centre for Supercomputing e.V.
(www.gauss-centre.eu) for funding this project by providing computing
time on the GCS Supercomputer SuperMUC at Leibniz Supercomputing Centre
(www.lrz.de).

\bibliographystyle{apsrev4-1modified}
\bibliography{biblio}

\begin{thebibliography}{64}%
\makeatletter
\providecommand \@ifxundefined [1]{%
 \@ifx{#1\undefined}
}%
\providecommand \@ifnum [1]{%
 \ifnum #1\expandafter \@firstoftwo
 \else \expandafter \@secondoftwo
 \fi
}%
\providecommand \@ifx [1]{%
 \ifx #1\expandafter \@firstoftwo
 \else \expandafter \@secondoftwo
 \fi
}%
\providecommand \natexlab [1]{#1}%
\providecommand \enquote  [1]{``#1''}%
\providecommand \bibnamefont  [1]{#1}%
\providecommand \bibfnamefont [1]{#1}%
\providecommand \citenamefont [1]{#1}%
\providecommand \href@noop [0]{\@secondoftwo}%
\providecommand \href [0]{\begingroup \@sanitize@url \@href}%
\providecommand \@href[1]{\@@startlink{#1}\@@href}%
\providecommand \@@href[1]{\endgroup#1\@@endlink}%
\providecommand \@sanitize@url [0]{\catcode `\\12\catcode `\$12\catcode
  `\&12\catcode `\#12\catcode `\^12\catcode `\_12\catcode `\%12\relax}%
\providecommand \@@startlink[1]{}%
\providecommand \@@endlink[0]{}%
\providecommand \url  [0]{\begingroup\@sanitize@url \@url }%
\providecommand \@url [1]{\endgroup\@href {#1}{\urlprefix }}%
\providecommand \urlprefix  [0]{URL }%
\providecommand \Eprint [0]{\href }%
\providecommand \doibase [0]{http://dx.doi.org/}%
\providecommand \selectlanguage [0]{\@gobble}%
\providecommand \bibinfo  [0]{\@secondoftwo}%
\providecommand \bibfield  [0]{\@secondoftwo}%
\providecommand \translation [1]{[#1]}%
\providecommand \BibitemOpen [0]{}%
\providecommand \bibitemStop [0]{}%
\providecommand \bibitemNoStop [0]{.\EOS\space}%
\providecommand \EOS [0]{\spacefactor3000\relax}%
\providecommand \BibitemShut  [1]{\csname bibitem#1\endcsname}%
\let\auto@bib@innerbib\@empty
\bibitem [{\citenamefont {Hasan}\ and\ \citenamefont {Kane}(2010)}]{RPMHasan}%
  \BibitemOpen
  \bibfield  {author} {\bibinfo {author} {\bibfnamefont {M.~Z.}\ \bibnamefont
  {Hasan}}\ and\ \bibinfo {author} {\bibfnamefont {C.~L.}\ \bibnamefont
  {Kane}},\ }\bibfield  {title} {\emph {\bibinfo {title} {{Colloquium:
  Topological insulators}},\ }}\href@noop {} {\bibfield  {journal} {\bibinfo
  {journal} {Rev. Mod. Phys.}\ }\textbf {\bibinfo {volume} {82}},\ \bibinfo
  {pages} {3045} (\bibinfo {year} {2010})}\BibitemShut {NoStop}%
\bibitem [{\citenamefont {Qi}\ and\ \citenamefont {Zhang}(2011)}]{RPMQi}%
  \BibitemOpen
  \bibfield  {author} {\bibinfo {author} {\bibfnamefont {X.-L.}\ \bibnamefont
  {Qi}}\ and\ \bibinfo {author} {\bibfnamefont {S.-C.}\ \bibnamefont {Zhang}},\
  }\bibfield  {title} {\emph {\bibinfo {title} {{Topological insulators and
  superconductors}},\ }}\href@noop {} {\bibfield  {journal} {\bibinfo
  {journal} {Rev. Mod. Phys.}\ }\textbf {\bibinfo {volume} {83}},\ \bibinfo
  {pages} {1057} (\bibinfo {year} {2011})}\BibitemShut {NoStop}%
\bibitem [{\citenamefont {Kane}\ and\ \citenamefont
  {Mele}(2005{\natexlab{a}})}]{KM1}%
  \BibitemOpen
  \bibfield  {author} {\bibinfo {author} {\bibfnamefont {C.~L.}\ \bibnamefont
  {Kane}}\ and\ \bibinfo {author} {\bibfnamefont {E.~J.}\ \bibnamefont
  {Mele}},\ }\bibfield  {title} {\emph {\bibinfo {title} {{${Z}_{2}$}},\
  }}\href@noop {} {\bibfield  {journal} {\bibinfo  {journal} {Phys. Rev.
  Lett.}\ }\textbf {\bibinfo {volume} {95}},\ \bibinfo {pages} {146802}
  (\bibinfo {year} {2005}{\natexlab{a}})}\BibitemShut {NoStop}%
\bibitem [{\citenamefont {Kane}\ and\ \citenamefont
  {Mele}(2005{\natexlab{b}})}]{KM2}%
  \BibitemOpen
  \bibfield  {author} {\bibinfo {author} {\bibfnamefont {C.~L.}\ \bibnamefont
  {Kane}}\ and\ \bibinfo {author} {\bibfnamefont {E.~J.}\ \bibnamefont
  {Mele}},\ }\bibfield  {title} {\emph {\bibinfo {title} {{Quantum Spin Hall
  Effect in Graphene}},\ }}\href@noop {} {\bibfield  {journal} {\bibinfo
  {journal} {Phys. Rev. Lett.}\ }\textbf {\bibinfo {volume} {95}},\ \bibinfo
  {pages} {226801} (\bibinfo {year} {2005}{\natexlab{b}})}\BibitemShut
  {NoStop}%
\bibitem [{\citenamefont {Bernevig}\ \emph {et~al.}(2006)\citenamefont
  {Bernevig}, \citenamefont {Hughes},\ and\ \citenamefont {Zhang}}]{Bernevig}%
  \BibitemOpen
  \bibfield  {author} {\bibinfo {author} {\bibfnamefont {B.~A.}\ \bibnamefont
  {Bernevig}}, \bibinfo {author} {\bibfnamefont {T.~L.}\ \bibnamefont
  {Hughes}}, \ and\ \bibinfo {author} {\bibfnamefont {S.-C.}\ \bibnamefont
  {Zhang}},\ }\bibfield  {title} {\emph {\bibinfo {title} {{Quantum Spin Hall
  Effect and Topological Phase Transition in HgTe Quantum Wells}},\
  }}\href@noop {} {\bibfield  {journal} {\bibinfo  {journal} {Science}\
  }\textbf {\bibinfo {volume} {314}},\ \bibinfo {pages} {1757} (\bibinfo {year}
  {2006})}\BibitemShut {NoStop}%
\bibitem [{\citenamefont {K{\"o}nig}\ \emph {et~al.}(2007)\citenamefont
  {K{\"o}nig}, \citenamefont {Wiedmann}, \citenamefont {Br{\"u}ne},
  \citenamefont {Roth}, \citenamefont {Buhmann}, \citenamefont {Molenkamp},
  \citenamefont {Qi},\ and\ \citenamefont {Zhang}}]{Konig}%
  \BibitemOpen
  \bibfield  {author} {\bibinfo {author} {\bibfnamefont {M.}~\bibnamefont
  {K{\"o}nig}}, \bibinfo {author} {\bibfnamefont {S.}~\bibnamefont {Wiedmann}},
  \bibinfo {author} {\bibfnamefont {C.}~\bibnamefont {Br{\"u}ne}}, \bibinfo
  {author} {\bibfnamefont {A.}~\bibnamefont {Roth}}, \bibinfo {author}
  {\bibfnamefont {H.}~\bibnamefont {Buhmann}}, \bibinfo {author} {\bibfnamefont
  {L.~W.}\ \bibnamefont {Molenkamp}}, \bibinfo {author} {\bibfnamefont {X.-L.}\
  \bibnamefont {Qi}}, \ and\ \bibinfo {author} {\bibfnamefont {S.-C.}\
  \bibnamefont {Zhang}},\ }\bibfield  {title} {\emph {\bibinfo {title}
  {{Quantum Spin Hall Insulator State in HgTe Quantum Wells}},\ }}\href@noop {}
  {\bibfield  {journal} {\bibinfo  {journal} {Science}\ }\textbf {\bibinfo
  {volume} {318}},\ \bibinfo {pages} {766} (\bibinfo {year}
  {2007})}\BibitemShut {NoStop}%
\bibitem [{\citenamefont {Molle}\ \emph {et~al.}(2017)\citenamefont {Molle},
  \citenamefont {Goldberger}, \citenamefont {Houssa}, \citenamefont {Xu},
  \citenamefont {Zhang},\ and\ \citenamefont {Akinwande}}]{2DXenes}%
  \BibitemOpen
  \bibfield  {author} {\bibinfo {author} {\bibfnamefont {A.}~\bibnamefont
  {Molle}}, \bibinfo {author} {\bibfnamefont {J.}~\bibnamefont {Goldberger}},
  \bibinfo {author} {\bibfnamefont {M.}~\bibnamefont {Houssa}}, \bibinfo
  {author} {\bibfnamefont {Y.}~\bibnamefont {Xu}}, \bibinfo {author}
  {\bibfnamefont {S.-C.}\ \bibnamefont {Zhang}}, \ and\ \bibinfo {author}
  {\bibfnamefont {D.}~\bibnamefont {Akinwande}},\ }\bibfield  {title} {\emph
  {\bibinfo {title} {{Buckled two-dimensional Xene sheets}},\ }}\href@noop {}
  {\bibfield  {journal} {\bibinfo  {journal} {Nature Materials}\ }\textbf
  {\bibinfo {volume} {16}},\ \bibinfo {pages} {163} (\bibinfo {year}
  {2017})}\BibitemShut {NoStop}%
\bibitem [{\citenamefont {Xu}\ \emph {et~al.}(2013)\citenamefont {Xu},
  \citenamefont {Yan}, \citenamefont {Zhang}, \citenamefont {Wang},
  \citenamefont {Xu}, \citenamefont {Tang}, \citenamefont {Duan},\ and\
  \citenamefont {Zhang}}]{Stanene}%
  \BibitemOpen
  \bibfield  {author} {\bibinfo {author} {\bibfnamefont {Y.}~\bibnamefont
  {Xu}}, \bibinfo {author} {\bibfnamefont {B.}~\bibnamefont {Yan}}, \bibinfo
  {author} {\bibfnamefont {H.-J.}\ \bibnamefont {Zhang}}, \bibinfo {author}
  {\bibfnamefont {J.}~\bibnamefont {Wang}}, \bibinfo {author} {\bibfnamefont
  {G.}~\bibnamefont {Xu}}, \bibinfo {author} {\bibfnamefont {P.}~\bibnamefont
  {Tang}}, \bibinfo {author} {\bibfnamefont {W.}~\bibnamefont {Duan}}, \ and\
  \bibinfo {author} {\bibfnamefont {S.-C.}\ \bibnamefont {Zhang}},\ }\bibfield
  {title} {\emph {\bibinfo {title} {{Large-Gap Quantum Spin Hall Insulators in
  Tin Films}},\ }}\href@noop {} {\bibfield  {journal} {\bibinfo  {journal}
  {Phys. Rev. Lett.}\ }\textbf {\bibinfo {volume} {111}},\ \bibinfo {pages}
  {136804} (\bibinfo {year} {2013})}\BibitemShut {NoStop}%
\bibitem [{\citenamefont {Geissler}\ \emph {et~al.}(2013)\citenamefont
  {Geissler}, \citenamefont {Budich},\ and\ \citenamefont
  {Trauzettel}}]{GeisslerNJP}%
  \BibitemOpen
  \bibfield  {author} {\bibinfo {author} {\bibfnamefont {F.}~\bibnamefont
  {Geissler}}, \bibinfo {author} {\bibfnamefont {J.~C.}\ \bibnamefont
  {Budich}}, \ and\ \bibinfo {author} {\bibfnamefont {B.}~\bibnamefont
  {Trauzettel}},\ }\bibfield  {title} {\emph {\bibinfo {title} {Group
  theoretical and topological analysis of the quantum spin hall effect in
  silicene},\ }}\href@noop {} {\bibfield  {journal} {\bibinfo  {journal} {New
  Journal of Physics}\ }\textbf {\bibinfo {volume} {15}},\ \bibinfo {pages}
  {085030} (\bibinfo {year} {2013})}\BibitemShut {NoStop}%
\bibitem [{\citenamefont {Zhang}\ \emph {et~al.}(2016)\citenamefont {Zhang},
  \citenamefont {Bampoulis}, \citenamefont {Rudenko}, \citenamefont {Yao},
  \citenamefont {van Houselt}, \citenamefont {Poelsema}, \citenamefont
  {Katsnelson},\ and\ \citenamefont {Zandvliet}}]{GermaneneMoS2}%
  \BibitemOpen
  \bibfield  {author} {\bibinfo {author} {\bibfnamefont {L.}~\bibnamefont
  {Zhang}}, \bibinfo {author} {\bibfnamefont {P.}~\bibnamefont {Bampoulis}},
  \bibinfo {author} {\bibfnamefont {A.~N.}\ \bibnamefont {Rudenko}}, \bibinfo
  {author} {\bibfnamefont {Q.}~\bibnamefont {Yao}}, \bibinfo {author}
  {\bibfnamefont {A.}~\bibnamefont {van Houselt}}, \bibinfo {author}
  {\bibfnamefont {B.}~\bibnamefont {Poelsema}}, \bibinfo {author}
  {\bibfnamefont {M.~I.}\ \bibnamefont {Katsnelson}}, \ and\ \bibinfo {author}
  {\bibfnamefont {H.~J.~W.}\ \bibnamefont {Zandvliet}},\ }\bibfield  {title}
  {\emph {\bibinfo {title} {{Structural and Electronic Properties of Germanene
  on ${\mathrm{MoS}}_{2}$}},\ }}\href@noop {} {\bibfield  {journal} {\bibinfo
  {journal} {Phys. Rev. Lett.}\ }\textbf {\bibinfo {volume} {116}},\ \bibinfo
  {pages} {256804} (\bibinfo {year} {2016})}\BibitemShut {NoStop}%
\bibitem [{\citenamefont {Zhu}\ \emph {et~al.}(2015)\citenamefont {Zhu},
  \citenamefont {Chen}, \citenamefont {Xu}, \citenamefont {Gao}, \citenamefont
  {Guan}, \citenamefont {Liu}, \citenamefont {Qian}, \citenamefont {Zhang},\
  and\ \citenamefont {Jia}}]{Zhu2015}%
  \BibitemOpen
  \bibfield  {author} {\bibinfo {author} {\bibfnamefont {F.-F.}\ \bibnamefont
  {Zhu}}, \bibinfo {author} {\bibfnamefont {W.-J.}\ \bibnamefont {Chen}},
  \bibinfo {author} {\bibfnamefont {Y.}~\bibnamefont {Xu}}, \bibinfo {author}
  {\bibfnamefont {C.-L.}\ \bibnamefont {Gao}}, \bibinfo {author} {\bibfnamefont
  {D.-D.}\ \bibnamefont {Guan}}, \bibinfo {author} {\bibfnamefont {C.-H.}\
  \bibnamefont {Liu}}, \bibinfo {author} {\bibfnamefont {D.}~\bibnamefont
  {Qian}}, \bibinfo {author} {\bibfnamefont {S.-C.}\ \bibnamefont {Zhang}}, \
  and\ \bibinfo {author} {\bibfnamefont {J.-F.}\ \bibnamefont {Jia}},\
  }\bibfield  {title} {\emph {\bibinfo {title} {{Epitaxial growth of
  two-dimensional stanene}},\ }}\href@noop {} {\bibfield  {journal} {\bibinfo
  {journal} {Nature Materials}\ }\textbf {\bibinfo {volume} {14}},\ \bibinfo
  {pages} {1020} (\bibinfo {year} {2015})}\BibitemShut {NoStop}%
\bibitem [{\citenamefont {Reis}\ \emph {et~al.}(2017)\citenamefont {Reis},
  \citenamefont {Li}, \citenamefont {Dudy}, \citenamefont {Bauernfeind},
  \citenamefont {Glass}, \citenamefont {Hanke}, \citenamefont {Thomale},
  \citenamefont {Sch{\"a}fer},\ and\ \citenamefont {Claessen}}]{Reis287}%
  \BibitemOpen
  \bibfield  {author} {\bibinfo {author} {\bibfnamefont {F.}~\bibnamefont
  {Reis}}, \bibinfo {author} {\bibfnamefont {G.}~\bibnamefont {Li}}, \bibinfo
  {author} {\bibfnamefont {L.}~\bibnamefont {Dudy}}, \bibinfo {author}
  {\bibfnamefont {M.}~\bibnamefont {Bauernfeind}}, \bibinfo {author}
  {\bibfnamefont {S.}~\bibnamefont {Glass}}, \bibinfo {author} {\bibfnamefont
  {W.}~\bibnamefont {Hanke}}, \bibinfo {author} {\bibfnamefont
  {R.}~\bibnamefont {Thomale}}, \bibinfo {author} {\bibfnamefont
  {J.}~\bibnamefont {Sch{\"a}fer}}, \ and\ \bibinfo {author} {\bibfnamefont
  {R.}~\bibnamefont {Claessen}},\ }\bibfield  {title} {\emph {\bibinfo {title}
  {{Bismuthene on a SiC substrate: A candidate for a high-temperature quantum
  spin Hall material}},\ }}\href@noop {} {\bibfield  {journal} {\bibinfo
  {journal} {Science}\ }\textbf {\bibinfo {volume} {357}},\ \bibinfo {pages}
  {287} (\bibinfo {year} {2017})}\BibitemShut {NoStop}%
\bibitem [{\citenamefont {Qian}\ \emph {et~al.}(2014)\citenamefont {Qian},
  \citenamefont {Liu}, \citenamefont {Fu},\ and\ \citenamefont {Li}}]{WTe2_1}%
  \BibitemOpen
  \bibfield  {author} {\bibinfo {author} {\bibfnamefont {X.}~\bibnamefont
  {Qian}}, \bibinfo {author} {\bibfnamefont {J.}~\bibnamefont {Liu}}, \bibinfo
  {author} {\bibfnamefont {L.}~\bibnamefont {Fu}}, \ and\ \bibinfo {author}
  {\bibfnamefont {J.}~\bibnamefont {Li}},\ }\bibfield  {title} {\emph {\bibinfo
  {title} {{Quantum spin Hall effect in two-dimensional transition metal
  dichalcogenides}},\ }}\href@noop {} {\bibfield  {journal} {\bibinfo
  {journal} {Science}\ }\textbf {\bibinfo {volume} {346}},\ \bibinfo {pages}
  {1344} (\bibinfo {year} {2014})}\BibitemShut {NoStop}%
\bibitem [{\citenamefont {Tang}\ \emph {et~al.}(2017)\citenamefont {Tang},
  \citenamefont {Zhang}, \citenamefont {Wong}, \citenamefont {Pedramrazi},
  \citenamefont {Tsai}, \citenamefont {Jia}, \citenamefont {Moritz},
  \citenamefont {Claassen}, \citenamefont {Ryu}, \citenamefont {Kahn},
  \citenamefont {Jiang}, \citenamefont {Yan}, \citenamefont {Hashimoto},
  \citenamefont {Lu}, \citenamefont {Moore}, \citenamefont {Hwang},
  \citenamefont {Hwang}, \citenamefont {Hussain}, \citenamefont {Chen},
  \citenamefont {Ugeda}, \citenamefont {Liu}, \citenamefont {Xie},
  \citenamefont {Devereaux}, \citenamefont {Crommie}, \citenamefont {Mo},\ and\
  \citenamefont {Shen}}]{WTe2_2}%
  \BibitemOpen
  \bibfield  {author} {\bibinfo {author} {\bibfnamefont {S.}~\bibnamefont
  {Tang}}, \bibinfo {author} {\bibfnamefont {C.}~\bibnamefont {Zhang}},
  \bibinfo {author} {\bibfnamefont {D.}~\bibnamefont {Wong}}, \bibinfo {author}
  {\bibfnamefont {Z.}~\bibnamefont {Pedramrazi}}, \bibinfo {author}
  {\bibfnamefont {H.-Z.}\ \bibnamefont {Tsai}}, \bibinfo {author}
  {\bibfnamefont {C.}~\bibnamefont {Jia}}, \bibinfo {author} {\bibfnamefont
  {B.}~\bibnamefont {Moritz}}, \bibinfo {author} {\bibfnamefont
  {M.}~\bibnamefont {Claassen}}, \bibinfo {author} {\bibfnamefont
  {H.}~\bibnamefont {Ryu}}, \bibinfo {author} {\bibfnamefont {S.}~\bibnamefont
  {Kahn}}, \bibinfo {author} {\bibfnamefont {J.}~\bibnamefont {Jiang}},
  \bibinfo {author} {\bibfnamefont {H.}~\bibnamefont {Yan}}, \bibinfo {author}
  {\bibfnamefont {M.}~\bibnamefont {Hashimoto}}, \bibinfo {author}
  {\bibfnamefont {D.}~\bibnamefont {Lu}}, \bibinfo {author} {\bibfnamefont
  {R.~G.}\ \bibnamefont {Moore}}, \bibinfo {author} {\bibfnamefont {C.-C.}\
  \bibnamefont {Hwang}}, \bibinfo {author} {\bibfnamefont {C.}~\bibnamefont
  {Hwang}}, \bibinfo {author} {\bibfnamefont {Z.}~\bibnamefont {Hussain}},
  \bibinfo {author} {\bibfnamefont {Y.}~\bibnamefont {Chen}}, \bibinfo {author}
  {\bibfnamefont {M.~M.}\ \bibnamefont {Ugeda}}, \bibinfo {author}
  {\bibfnamefont {Z.}~\bibnamefont {Liu}}, \bibinfo {author} {\bibfnamefont
  {X.}~\bibnamefont {Xie}}, \bibinfo {author} {\bibfnamefont {T.~P.}\
  \bibnamefont {Devereaux}}, \bibinfo {author} {\bibfnamefont {M.~F.}\
  \bibnamefont {Crommie}}, \bibinfo {author} {\bibfnamefont {S.-K.}\
  \bibnamefont {Mo}}, \ and\ \bibinfo {author} {\bibfnamefont {Z.-X.}\
  \bibnamefont {Shen}},\ }\bibfield  {title} {\emph {\bibinfo {title} {{Quantum
  spin Hall state in monolayer WTe$_2$}},\ }}\href@noop {} {\bibfield
  {journal} {\bibinfo  {journal} {Nat. Phys.}\ }\textbf {\bibinfo {volume}
  {13}},\ \bibinfo {pages} {683} (\bibinfo {year} {2017})}\BibitemShut
  {NoStop}%
\bibitem [{\citenamefont {{Ok}}\ \emph {et~al.}()\citenamefont {{Ok}},
  \citenamefont {{Muechler}}, \citenamefont {{Di Sante}}, \citenamefont
  {{Sangiovanni}}, \citenamefont {{Thomale}},\ and\ \citenamefont
  {{Neupert}}}]{WTe2}%
  \BibitemOpen
  \bibfield  {author} {\bibinfo {author} {\bibfnamefont {S.}~\bibnamefont
  {{Ok}}}, \bibinfo {author} {\bibfnamefont {L.}~\bibnamefont {{Muechler}}},
  \bibinfo {author} {\bibfnamefont {D.}~\bibnamefont {{Di Sante}}}, \bibinfo
  {author} {\bibfnamefont {G.}~\bibnamefont {{Sangiovanni}}}, \bibinfo {author}
  {\bibfnamefont {R.}~\bibnamefont {{Thomale}}}, \ and\ \bibinfo {author}
  {\bibfnamefont {T.}~\bibnamefont {{Neupert}}},\ }\bibfield  {title} {\emph
  {\bibinfo {title} {{Custodial glide symmetry of quantum spin Hall edge modes
  in WTe{$_2$} monolayer}},\ }}\href@noop {} {\bibinfo  {journal}
  {arXiv:1811.00551}\ }\BibitemShut {NoStop}%
\bibitem [{\citenamefont {Xu}\ \emph {et~al.}(2018)\citenamefont {Xu},
  \citenamefont {Chan}, \citenamefont {Chen}, \citenamefont {Wang},
  \citenamefont {Fl\"ototto}, \citenamefont {Hlevyack}, \citenamefont {Bian},
  \citenamefont {Mo}, \citenamefont {Chou},\ and\ \citenamefont
  {Chiang}}]{StaneneInSb}%
  \BibitemOpen
\bibfield  {journal} {  }\bibfield  {author} {\bibinfo {author} {\bibfnamefont
  {C.-Z.}\ \bibnamefont {Xu}}, \bibinfo {author} {\bibfnamefont {Y.-H.}\
  \bibnamefont {Chan}}, \bibinfo {author} {\bibfnamefont {P.}~\bibnamefont
  {Chen}}, \bibinfo {author} {\bibfnamefont {X.}~\bibnamefont {Wang}}, \bibinfo
  {author} {\bibfnamefont {D.}~\bibnamefont {Fl\"ototto}}, \bibinfo {author}
  {\bibfnamefont {J.~A.}\ \bibnamefont {Hlevyack}}, \bibinfo {author}
  {\bibfnamefont {G.}~\bibnamefont {Bian}}, \bibinfo {author} {\bibfnamefont
  {S.-K.}\ \bibnamefont {Mo}}, \bibinfo {author} {\bibfnamefont {M.-Y.}\
  \bibnamefont {Chou}}, \ and\ \bibinfo {author} {\bibfnamefont {T.-C.}\
  \bibnamefont {Chiang}},\ }\bibfield  {title} {\emph {\bibinfo {title}
  {{Gapped electronic structure of epitaxial stanene on InSb(111)}},\
  }}\href@noop {} {\bibfield  {journal} {\bibinfo  {journal} {Phys. Rev. B}\
  }\textbf {\bibinfo {volume} {97}},\ \bibinfo {pages} {035122} (\bibinfo
  {year} {2018})}\BibitemShut {NoStop}%
\bibitem [{\citenamefont {Ni}\ \emph {et~al.}(2017)\citenamefont {Ni},
  \citenamefont {Minamitani}, \citenamefont {Ando},\ and\ \citenamefont
  {Watanabe}}]{ZeyuanNi}%
  \BibitemOpen
  \bibfield  {author} {\bibinfo {author} {\bibfnamefont {Z.}~\bibnamefont
  {Ni}}, \bibinfo {author} {\bibfnamefont {E.}~\bibnamefont {Minamitani}},
  \bibinfo {author} {\bibfnamefont {Y.}~\bibnamefont {Ando}}, \ and\ \bibinfo
  {author} {\bibfnamefont {S.}~\bibnamefont {Watanabe}},\ }\bibfield  {title}
  {\emph {\bibinfo {title} {{Germanene and stanene on two-dimensional
  substrates: Dirac cone and ${Z}_{2}$ invariant}},\ }}\href@noop {} {\bibfield
   {journal} {\bibinfo  {journal} {Phys. Rev. B}\ }\textbf {\bibinfo {volume}
  {96}},\ \bibinfo {pages} {075427} (\bibinfo {year} {2017})}\BibitemShut
  {NoStop}%
\bibitem [{\citenamefont {Matusalem}\ \emph {et~al.}(2016)\citenamefont
  {Matusalem}, \citenamefont {Bechstedt}, \citenamefont {Marques},\ and\
  \citenamefont {Teles}}]{Bechstedt}%
  \BibitemOpen
  \bibfield  {author} {\bibinfo {author} {\bibfnamefont {F.}~\bibnamefont
  {Matusalem}}, \bibinfo {author} {\bibfnamefont {F.}~\bibnamefont
  {Bechstedt}}, \bibinfo {author} {\bibfnamefont {M.}~\bibnamefont {Marques}},
  \ and\ \bibinfo {author} {\bibfnamefont {L.~K.}\ \bibnamefont {Teles}},\
  }\bibfield  {title} {\emph {\bibinfo {title} {{Quantum spin Hall phase in
  stanene-derived overlayers on passivated SiC substrates}},\ }}\href@noop {}
  {\bibfield  {journal} {\bibinfo  {journal} {Phys. Rev. B}\ }\textbf {\bibinfo
  {volume} {94}},\ \bibinfo {pages} {241403} (\bibinfo {year}
  {2016})}\BibitemShut {NoStop}%
\bibitem [{\citenamefont {Matusalem}\ \emph {et~al.}(2017)\citenamefont
  {Matusalem}, \citenamefont {Koda}, \citenamefont {Bechstedt}, \citenamefont
  {Marques},\ and\ \citenamefont {Teles}}]{Matusalem}%
  \BibitemOpen
  \bibfield  {author} {\bibinfo {author} {\bibfnamefont {F.}~\bibnamefont
  {Matusalem}}, \bibinfo {author} {\bibfnamefont {D.~S.}\ \bibnamefont {Koda}},
  \bibinfo {author} {\bibfnamefont {F.}~\bibnamefont {Bechstedt}}, \bibinfo
  {author} {\bibfnamefont {M.}~\bibnamefont {Marques}}, \ and\ \bibinfo
  {author} {\bibfnamefont {L.~K.}\ \bibnamefont {Teles}},\ }\bibfield  {title}
  {\emph {\bibinfo {title} {{Deposition of topological silicene, germanene and
  stanene on graphene-covered ${Si}{C}$ substrates}},\ }}\href@noop {}
  {\bibfield  {journal} {\bibinfo  {journal} {Scientific Reports}\ }\textbf
  {\bibinfo {volume} {7}},\ \bibinfo {pages} {15700} (\bibinfo {year}
  {2017})}\BibitemShut {NoStop}%
\bibitem [{\citenamefont {Glass}\ \emph {et~al.}(2015)\citenamefont {Glass},
  \citenamefont {Li}, \citenamefont {Adler}, \citenamefont {Aulbach},
  \citenamefont {Fleszar}, \citenamefont {Thomale}, \citenamefont {Hanke},
  \citenamefont {Claessen},\ and\ \citenamefont {Sch\"afer}}]{PRLJoerg}%
  \BibitemOpen
  \bibfield  {author} {\bibinfo {author} {\bibfnamefont {S.}~\bibnamefont
  {Glass}}, \bibinfo {author} {\bibfnamefont {G.}~\bibnamefont {Li}}, \bibinfo
  {author} {\bibfnamefont {F.}~\bibnamefont {Adler}}, \bibinfo {author}
  {\bibfnamefont {J.}~\bibnamefont {Aulbach}}, \bibinfo {author} {\bibfnamefont
  {A.}~\bibnamefont {Fleszar}}, \bibinfo {author} {\bibfnamefont
  {R.}~\bibnamefont {Thomale}}, \bibinfo {author} {\bibfnamefont
  {W.}~\bibnamefont {Hanke}}, \bibinfo {author} {\bibfnamefont
  {R.}~\bibnamefont {Claessen}}, \ and\ \bibinfo {author} {\bibfnamefont
  {J.}~\bibnamefont {Sch\"afer}},\ }\bibfield  {title} {\emph {\bibinfo {title}
  {{Triangular Spin-Orbit-Coupled Lattice with Strong Coulomb Correlations: Sn
  Atoms on a SiC(0001) Substrate}},\ }}\href@noop {} {\bibfield  {journal}
  {\bibinfo  {journal} {Phys. Rev. Lett.}\ }\textbf {\bibinfo {volume} {114}},\
  \bibinfo {pages} {247602} (\bibinfo {year} {2015})}\BibitemShut {NoStop}%
\bibitem [{\citenamefont {Glass}\ \emph {et~al.}(2016)\citenamefont {Glass},
  \citenamefont {Reis}, \citenamefont {Bauernfeind}, \citenamefont {Aulbach},
  \citenamefont {Scholz}, \citenamefont {Adler}, \citenamefont {Dudy},
  \citenamefont {Li}, \citenamefont {Claessen},\ and\ \citenamefont
  {Sch\"afer}}]{Glass2016}%
  \BibitemOpen
  \bibfield  {author} {\bibinfo {author} {\bibfnamefont {S.}~\bibnamefont
  {Glass}}, \bibinfo {author} {\bibfnamefont {F.}~\bibnamefont {Reis}},
  \bibinfo {author} {\bibfnamefont {M.}~\bibnamefont {Bauernfeind}}, \bibinfo
  {author} {\bibfnamefont {J.}~\bibnamefont {Aulbach}}, \bibinfo {author}
  {\bibfnamefont {M.~R.}\ \bibnamefont {Scholz}}, \bibinfo {author}
  {\bibfnamefont {F.}~\bibnamefont {Adler}}, \bibinfo {author} {\bibfnamefont
  {L.}~\bibnamefont {Dudy}}, \bibinfo {author} {\bibfnamefont {G.}~\bibnamefont
  {Li}}, \bibinfo {author} {\bibfnamefont {R.}~\bibnamefont {Claessen}}, \ and\
  \bibinfo {author} {\bibfnamefont {J.}~\bibnamefont {Sch\"afer}},\ }\bibfield
  {title} {\emph {\bibinfo {title} {{Atomic-Scale Mapping of Layer-by-Layer
  Hydrogen Etching and Passivation of ${Si}{C}$(0001) Substrates}},\
  }}\href@noop {} {\bibfield  {journal} {\bibinfo  {journal} {The Journal of
  Physical Chemistry C}\ }\textbf {\bibinfo {volume} {120}},\ \bibinfo {pages}
  {10361} (\bibinfo {year} {2016})}\BibitemShut {NoStop}%
\bibitem [{\citenamefont {Houssa}\ \emph {et~al.}(2010)\citenamefont {Houssa},
  \citenamefont {Pourtois}, \citenamefont {V.}, \citenamefont {Afanas\'{e}v},\
  and\ \citenamefont {Stesmans}}]{houssa2010}%
  \BibitemOpen
  \bibfield  {author} {\bibinfo {author} {\bibfnamefont {M.}~\bibnamefont
  {Houssa}}, \bibinfo {author} {\bibfnamefont {G.}~\bibnamefont {Pourtois}},
  \bibinfo {author} {\bibfnamefont {V.}~\bibnamefont {V.}}, \bibinfo {author}
  {\bibnamefont {Afanas\'{e}v}}, \ and\ \bibinfo {author} {\bibfnamefont
  {A.}~\bibnamefont {Stesmans}},\ }\bibfield  {title} {\emph {\bibinfo {title}
  {Can silicon behave like graphene? a first-principles study},\ }}\href@noop
  {} {\bibfield  {journal} {\bibinfo  {journal} {Applied Physics Letters}\
  }\textbf {\bibinfo {volume} {97}},\ \bibinfo {pages} {112106} (\bibinfo
  {year} {2010})}\BibitemShut {NoStop}%
\bibitem [{\citenamefont {Kanno}\ \emph {et~al.}(2014)\citenamefont {Kanno},
  \citenamefont {Arafune}, \citenamefont {Lin}, \citenamefont {Minamitani},
  \citenamefont {Kawai},\ and\ \citenamefont {Takagi}}]{kanno2014}%
  \BibitemOpen
  \bibfield  {author} {\bibinfo {author} {\bibfnamefont {M.}~\bibnamefont
  {Kanno}}, \bibinfo {author} {\bibfnamefont {R.}~\bibnamefont {Arafune}},
  \bibinfo {author} {\bibfnamefont {C.~L.}\ \bibnamefont {Lin}}, \bibinfo
  {author} {\bibfnamefont {E.}~\bibnamefont {Minamitani}}, \bibinfo {author}
  {\bibfnamefont {M.}~\bibnamefont {Kawai}}, \ and\ \bibinfo {author}
  {\bibfnamefont {N.}~\bibnamefont {Takagi}},\ }\bibfield  {title} {\emph
  {\bibinfo {title} {Electronic decoupling by h-${B}{N}$ layer between silicene
  and ${Cu}$(111): A dft-based analysis},\ }}\href@noop {} {\bibfield
  {journal} {\bibinfo  {journal} {New Journal of Physics}\ }\textbf {\bibinfo
  {volume} {16}},\ \bibinfo {pages} {105019} (\bibinfo {year}
  {2014})}\BibitemShut {NoStop}%
\bibitem [{\citenamefont {d'Acapito}\ \emph {et~al.}(2016)\citenamefont
  {d'Acapito}, \citenamefont {Torrengo}, \citenamefont {Xenogiannopoulou},
  \citenamefont {Tsipas}, \citenamefont {Velasco}, \citenamefont {Tsoutsou},\
  and\ \citenamefont {Dimoulas}}]{dacapito2016}%
  \BibitemOpen
  \bibfield  {author} {\bibinfo {author} {\bibfnamefont {F.}~\bibnamefont
  {d'Acapito}}, \bibinfo {author} {\bibfnamefont {S.}~\bibnamefont {Torrengo}},
  \bibinfo {author} {\bibfnamefont {E.}~\bibnamefont {Xenogiannopoulou}},
  \bibinfo {author} {\bibfnamefont {P.}~\bibnamefont {Tsipas}}, \bibinfo
  {author} {\bibfnamefont {J.~M.}\ \bibnamefont {Velasco}}, \bibinfo {author}
  {\bibfnamefont {D.}~\bibnamefont {Tsoutsou}}, \ and\ \bibinfo {author}
  {\bibfnamefont {A.}~\bibnamefont {Dimoulas}},\ }\bibfield  {title} {\emph
  {\bibinfo {title} {{Evidence for Germanene growth on epitaxial hexagonal
  (h)-${Al}{N}$ on ${Ag}$(111)}},\ }}\href@noop {} {\bibfield  {journal}
  {\bibinfo  {journal} {Journal of Physics: Condensed Matter}\ }\textbf
  {\bibinfo {volume} {28}},\ \bibinfo {pages} {045002} (\bibinfo {year}
  {2016})}\BibitemShut {NoStop}%
\bibitem [{\citenamefont {Ni}\ \emph {et~al.}(2012)\citenamefont {Ni},
  \citenamefont {Liu}, \citenamefont {Tang}, \citenamefont {Zheng},
  \citenamefont {Zhou}, \citenamefont {Qin}, \citenamefont {Gao}, \citenamefont
  {Yu},\ and\ \citenamefont {Lu}}]{niNanoLett12}%
  \BibitemOpen
  \bibfield  {author} {\bibinfo {author} {\bibfnamefont {Z.}~\bibnamefont
  {Ni}}, \bibinfo {author} {\bibfnamefont {Q.}~\bibnamefont {Liu}}, \bibinfo
  {author} {\bibfnamefont {K.}~\bibnamefont {Tang}}, \bibinfo {author}
  {\bibfnamefont {J.}~\bibnamefont {Zheng}}, \bibinfo {author} {\bibfnamefont
  {J.}~\bibnamefont {Zhou}}, \bibinfo {author} {\bibfnamefont {R.}~\bibnamefont
  {Qin}}, \bibinfo {author} {\bibfnamefont {Z.}~\bibnamefont {Gao}}, \bibinfo
  {author} {\bibfnamefont {D.}~\bibnamefont {Yu}}, \ and\ \bibinfo {author}
  {\bibfnamefont {J.}~\bibnamefont {Lu}},\ }\bibfield  {title} {\emph {\bibinfo
  {title} {Tunable bandgap in silicene and germanene},\ }}\href@noop {}
  {\bibfield  {journal} {\bibinfo  {journal} {Nano Letters}\ }\textbf {\bibinfo
  {volume} {12}},\ \bibinfo {pages} {113} (\bibinfo {year} {2012})}\BibitemShut
  {NoStop}%
\bibitem [{\citenamefont {Du}\ \emph {et~al.}(2016)\citenamefont {Du},
  \citenamefont {Zhuang}, \citenamefont {Wang}, \citenamefont {Li},
  \citenamefont {Liu}, \citenamefont {Zhao}, \citenamefont {Xu}, \citenamefont
  {Feng}, \citenamefont {Chen}, \citenamefont {Wu}, \citenamefont {Wang},\ and\
  \citenamefont {Dou}}]{dueSciAdv2016}%
  \BibitemOpen
  \bibfield  {author} {\bibinfo {author} {\bibfnamefont {Y.}~\bibnamefont
  {Du}}, \bibinfo {author} {\bibfnamefont {J.}~\bibnamefont {Zhuang}}, \bibinfo
  {author} {\bibfnamefont {J.}~\bibnamefont {Wang}}, \bibinfo {author}
  {\bibfnamefont {Z.}~\bibnamefont {Li}}, \bibinfo {author} {\bibfnamefont
  {H.}~\bibnamefont {Liu}}, \bibinfo {author} {\bibfnamefont {J.}~\bibnamefont
  {Zhao}}, \bibinfo {author} {\bibfnamefont {X.}~\bibnamefont {Xu}}, \bibinfo
  {author} {\bibfnamefont {H.}~\bibnamefont {Feng}}, \bibinfo {author}
  {\bibfnamefont {L.}~\bibnamefont {Chen}}, \bibinfo {author} {\bibfnamefont
  {K.}~\bibnamefont {Wu}}, \bibinfo {author} {\bibfnamefont {X.}~\bibnamefont
  {Wang}}, \ and\ \bibinfo {author} {\bibfnamefont {S.~X.}\ \bibnamefont
  {Dou}},\ }\bibfield  {title} {\emph {\bibinfo {title} {Quasi-freestanding
  epitaxial silicene on ${Ag}$(111) by oxygen intercalation},\ }}\href@noop {}
  {\bibfield  {journal} {\bibinfo  {journal} {Science Advances}\ }\textbf
  {\bibinfo {volume} {2}} (\bibinfo {year} {2016})}\BibitemShut {NoStop}%
\bibitem [{\citenamefont {Kaloni}\ and\ \citenamefont
  {Schwingenschl\"ogl}(2013)}]{kaloni2013}%
  \BibitemOpen
  \bibfield  {author} {\bibinfo {author} {\bibfnamefont {T.~P.}\ \bibnamefont
  {Kaloni}}\ and\ \bibinfo {author} {\bibfnamefont {U.}~\bibnamefont
  {Schwingenschl\"ogl}},\ }\bibfield  {title} {\emph {\bibinfo {title} {Weak
  interaction between germanene and ${Ga}{As}$(0001) by h intercalation: A
  route to exfoliation},\ }}\href@noop {} {\bibfield  {journal} {\bibinfo
  {journal} {Journal of Applied Physics}\ }\textbf {\bibinfo {volume} {114}},\
  \bibinfo {pages} {184307} (\bibinfo {year} {2013})}\BibitemShut {NoStop}%
\bibitem [{\citenamefont {Virojanadara}\ \emph {et~al.}(2010)\citenamefont
  {Virojanadara}, \citenamefont {Watcharinyanon}, \citenamefont {Zakharov},\
  and\ \citenamefont {Johansson}}]{virojanadaraPRB82}%
  \BibitemOpen
  \bibfield  {author} {\bibinfo {author} {\bibfnamefont {C.}~\bibnamefont
  {Virojanadara}}, \bibinfo {author} {\bibfnamefont {S.}~\bibnamefont
  {Watcharinyanon}}, \bibinfo {author} {\bibfnamefont {A.~A.}\ \bibnamefont
  {Zakharov}}, \ and\ \bibinfo {author} {\bibfnamefont {L.~I.}\ \bibnamefont
  {Johansson}},\ }\bibfield  {title} {\emph {\bibinfo {title} {{Epitaxial
  graphene on 6${H}-{Si}{C}$ and ${Li}$ intercalation}},\ }}\href@noop {}
  {\bibfield  {journal} {\bibinfo  {journal} {Phys. Rev. B}\ }\textbf {\bibinfo
  {volume} {82}},\ \bibinfo {pages} {205402} (\bibinfo {year}
  {2010})}\BibitemShut {NoStop}%
\bibitem [{\citenamefont {Enderlein}\ \emph {et~al.}(2010)\citenamefont
  {Enderlein}, \citenamefont {Kim}, \citenamefont {Bostwick}, \citenamefont
  {Rotenberg},\ and\ \citenamefont {Horn}}]{enderleinNJP12}%
  \BibitemOpen
  \bibfield  {author} {\bibinfo {author} {\bibfnamefont {C.}~\bibnamefont
  {Enderlein}}, \bibinfo {author} {\bibfnamefont {Y.~S.}\ \bibnamefont {Kim}},
  \bibinfo {author} {\bibfnamefont {A.}~\bibnamefont {Bostwick}}, \bibinfo
  {author} {\bibfnamefont {E.}~\bibnamefont {Rotenberg}}, \ and\ \bibinfo
  {author} {\bibfnamefont {K.}~\bibnamefont {Horn}},\ }\bibfield  {title}
  {\emph {\bibinfo {title} {The formation of an energy gap in graphene on
  ruthenium by controlling the interface},\ }}\href@noop {} {\bibfield
  {journal} {\bibinfo  {journal} {New Journal of Physics}\ }\textbf {\bibinfo
  {volume} {12}},\ \bibinfo {pages} {033014} (\bibinfo {year}
  {2010})}\BibitemShut {NoStop}%
\bibitem [{\citenamefont {Oida}\ \emph {et~al.}(2010)\citenamefont {Oida},
  \citenamefont {McFeely}, \citenamefont {Hannon}, \citenamefont {Tromp},
  \citenamefont {Copel}, \citenamefont {Chen}, \citenamefont {Sun},
  \citenamefont {Farmer},\ and\ \citenamefont {Yurkas}}]{oidaPRB82}%
  \BibitemOpen
  \bibfield  {author} {\bibinfo {author} {\bibfnamefont {S.}~\bibnamefont
  {Oida}}, \bibinfo {author} {\bibfnamefont {F.~R.}\ \bibnamefont {McFeely}},
  \bibinfo {author} {\bibfnamefont {J.~B.}\ \bibnamefont {Hannon}}, \bibinfo
  {author} {\bibfnamefont {R.~M.}\ \bibnamefont {Tromp}}, \bibinfo {author}
  {\bibfnamefont {M.}~\bibnamefont {Copel}}, \bibinfo {author} {\bibfnamefont
  {Z.}~\bibnamefont {Chen}}, \bibinfo {author} {\bibfnamefont {Y.}~\bibnamefont
  {Sun}}, \bibinfo {author} {\bibfnamefont {D.~B.}\ \bibnamefont {Farmer}}, \
  and\ \bibinfo {author} {\bibfnamefont {J.}~\bibnamefont {Yurkas}},\
  }\bibfield  {title} {\emph {\bibinfo {title} {Decoupling graphene from
  ${Si}{C}$(0001) via oxidation},\ }}\href@noop {} {\bibfield  {journal}
  {\bibinfo  {journal} {Phys. Rev. B}\ }\textbf {\bibinfo {volume} {82}},\
  \bibinfo {pages} {041411} (\bibinfo {year} {2010})}\BibitemShut {NoStop}%
\bibitem [{\citenamefont {Varchon}\ \emph {et~al.}(2007)\citenamefont
  {Varchon}, \citenamefont {Feng}, \citenamefont {Hass}, \citenamefont {Li},
  \citenamefont {Nguyen}, \citenamefont {Naud}, \citenamefont {Mallet},
  \citenamefont {Veuillen}, \citenamefont {Berger}, \citenamefont {Conrad},\
  and\ \citenamefont {Magaud}}]{varchonPRL99}%
  \BibitemOpen
  \bibfield  {author} {\bibinfo {author} {\bibfnamefont {F.}~\bibnamefont
  {Varchon}}, \bibinfo {author} {\bibfnamefont {R.}~\bibnamefont {Feng}},
  \bibinfo {author} {\bibfnamefont {J.}~\bibnamefont {Hass}}, \bibinfo {author}
  {\bibfnamefont {X.}~\bibnamefont {Li}}, \bibinfo {author} {\bibfnamefont
  {B.~N.}\ \bibnamefont {Nguyen}}, \bibinfo {author} {\bibfnamefont
  {C.}~\bibnamefont {Naud}}, \bibinfo {author} {\bibfnamefont {P.}~\bibnamefont
  {Mallet}}, \bibinfo {author} {\bibfnamefont {J.-Y.}\ \bibnamefont
  {Veuillen}}, \bibinfo {author} {\bibfnamefont {C.}~\bibnamefont {Berger}},
  \bibinfo {author} {\bibfnamefont {E.~H.}\ \bibnamefont {Conrad}}, \ and\
  \bibinfo {author} {\bibfnamefont {L.}~\bibnamefont {Magaud}},\ }\bibfield
  {title} {\emph {\bibinfo {title} {Electronic structure of epitaxial graphene
  layers on ${Si}{C}$: Effect of the substrate},\ }}\href@noop {} {\bibfield
  {journal} {\bibinfo  {journal} {Phys. Rev. Lett.}\ }\textbf {\bibinfo
  {volume} {99}},\ \bibinfo {pages} {126805} (\bibinfo {year}
  {2007})}\BibitemShut {NoStop}%
\bibitem [{\citenamefont {Watcharinyanon}\ \emph {et~al.}(2011)\citenamefont
  {Watcharinyanon}, \citenamefont {Virojanadara}, \citenamefont {Osiecki},
  \citenamefont {Zakharov}, \citenamefont {Yakimova}, \citenamefont {Uhrberg},\
  and\ \citenamefont {Johansson}}]{watcharinyanonSurfSci2011}%
  \BibitemOpen
  \bibfield  {author} {\bibinfo {author} {\bibfnamefont {S.}~\bibnamefont
  {Watcharinyanon}}, \bibinfo {author} {\bibfnamefont {C.}~\bibnamefont
  {Virojanadara}}, \bibinfo {author} {\bibfnamefont {J.}~\bibnamefont
  {Osiecki}}, \bibinfo {author} {\bibfnamefont {A.}~\bibnamefont {Zakharov}},
  \bibinfo {author} {\bibfnamefont {R.}~\bibnamefont {Yakimova}}, \bibinfo
  {author} {\bibfnamefont {R.}~\bibnamefont {Uhrberg}}, \ and\ \bibinfo
  {author} {\bibfnamefont {L.}~\bibnamefont {Johansson}},\ }\bibfield  {title}
  {\emph {\bibinfo {title} {Hydrogen intercalation of graphene grown on
  6${H}-{Si}{C}$(0001)},\ }}\href@noop {} {\bibfield  {journal} {\bibinfo
  {journal} {Surface Science}\ }\textbf {\bibinfo {volume} {605}},\ \bibinfo
  {pages} {1662 } (\bibinfo {year} {2011})}\BibitemShut {NoStop}%
\bibitem [{\citenamefont {Emtsev}\ \emph {et~al.}(2009)\citenamefont {Emtsev},
  \citenamefont {Bostwick}, \citenamefont {Horn}, \citenamefont {Jobst},
  \citenamefont {Kellogg}, \citenamefont {Ley}, \citenamefont {McChesney},
  \citenamefont {Ohta}, \citenamefont {Reshanov}, \citenamefont {R{\"o}hrl},
  \citenamefont {Rotenberg}, \citenamefont {Schmid}, \citenamefont {Waldmann},
  \citenamefont {Weber},\ and\ \citenamefont {Seyller}}]{Emtsev2009}%
  \BibitemOpen
  \bibfield  {author} {\bibinfo {author} {\bibfnamefont {K.~V.}\ \bibnamefont
  {Emtsev}}, \bibinfo {author} {\bibfnamefont {A.}~\bibnamefont {Bostwick}},
  \bibinfo {author} {\bibfnamefont {K.}~\bibnamefont {Horn}}, \bibinfo {author}
  {\bibfnamefont {J.}~\bibnamefont {Jobst}}, \bibinfo {author} {\bibfnamefont
  {G.~L.}\ \bibnamefont {Kellogg}}, \bibinfo {author} {\bibfnamefont
  {L.}~\bibnamefont {Ley}}, \bibinfo {author} {\bibfnamefont {J.~L.}\
  \bibnamefont {McChesney}}, \bibinfo {author} {\bibfnamefont {T.}~\bibnamefont
  {Ohta}}, \bibinfo {author} {\bibfnamefont {S.~A.}\ \bibnamefont {Reshanov}},
  \bibinfo {author} {\bibfnamefont {J.}~\bibnamefont {R{\"o}hrl}}, \bibinfo
  {author} {\bibfnamefont {E.}~\bibnamefont {Rotenberg}}, \bibinfo {author}
  {\bibfnamefont {A.~K.}\ \bibnamefont {Schmid}}, \bibinfo {author}
  {\bibfnamefont {D.}~\bibnamefont {Waldmann}}, \bibinfo {author}
  {\bibfnamefont {H.~B.}\ \bibnamefont {Weber}}, \ and\ \bibinfo {author}
  {\bibfnamefont {T.}~\bibnamefont {Seyller}},\ }\bibfield  {title} {\emph
  {\bibinfo {title} {Towards wafer-size graphene layers by atmospheric pressure
  graphitization of silicon carbide},\ }}\href@noop {} {\bibfield  {journal}
  {\bibinfo  {journal} {Nature Materials}\ }\textbf {\bibinfo {volume} {8}},\
  \bibinfo {pages} {203} (\bibinfo {year} {2009})}\BibitemShut {NoStop}%
\bibitem [{\citenamefont {Zhou}\ \emph {et~al.}(2007)\citenamefont {Zhou},
  \citenamefont {Gweon}, \citenamefont {Fedorov}, \citenamefont {First},
  \citenamefont {de~Heer}, \citenamefont {Lee}, \citenamefont {Guinea},
  \citenamefont {Castro~Neto},\ and\ \citenamefont {Lanzara}}]{Zhou2007}%
  \BibitemOpen
  \bibfield  {author} {\bibinfo {author} {\bibfnamefont {S.~Y.}\ \bibnamefont
  {Zhou}}, \bibinfo {author} {\bibfnamefont {G.-H.}\ \bibnamefont {Gweon}},
  \bibinfo {author} {\bibfnamefont {A.~V.}\ \bibnamefont {Fedorov}}, \bibinfo
  {author} {\bibfnamefont {P.~N.}\ \bibnamefont {First}}, \bibinfo {author}
  {\bibfnamefont {W.~A.}\ \bibnamefont {de~Heer}}, \bibinfo {author}
  {\bibfnamefont {D.-H.}\ \bibnamefont {Lee}}, \bibinfo {author} {\bibfnamefont
  {F.}~\bibnamefont {Guinea}}, \bibinfo {author} {\bibfnamefont {A.~H.}\
  \bibnamefont {Castro~Neto}}, \ and\ \bibinfo {author} {\bibfnamefont
  {A.}~\bibnamefont {Lanzara}},\ }\bibfield  {title} {\emph {\bibinfo {title}
  {Substrate-induced bandgap opening in epitaxial graphene},\ }}\href@noop {}
  {\bibfield  {journal} {\bibinfo  {journal} {Nature Materials}\ }\textbf
  {\bibinfo {volume} {6}},\ \bibinfo {pages} {770} (\bibinfo {year}
  {2007})}\BibitemShut {NoStop}%
\bibitem [{\citenamefont {Semenoff}(1984)}]{Semenoff}%
  \BibitemOpen
  \bibfield  {author} {\bibinfo {author} {\bibfnamefont {G.~W.}\ \bibnamefont
  {Semenoff}},\ }\bibfield  {title} {\emph {\bibinfo {title} {{Condensed-Matter
  Simulation of a Three-Dimensional Anomaly}},\ }}\href@noop {} {\bibfield
  {journal} {\bibinfo  {journal} {Phys. Rev. Lett.}\ }\textbf {\bibinfo
  {volume} {53}},\ \bibinfo {pages} {2449} (\bibinfo {year}
  {1984})}\BibitemShut {NoStop}%
\bibitem [{\citenamefont {Kresse}\ and\ \citenamefont
  {Furthm\"{u}ller}(1996)}]{VASP1}%
  \BibitemOpen
  \bibfield  {author} {\bibinfo {author} {\bibfnamefont {G.}~\bibnamefont
  {Kresse}}\ and\ \bibinfo {author} {\bibfnamefont {J.}~\bibnamefont
  {Furthm\"{u}ller}},\ }\bibfield  {title} {\emph {\bibinfo {title} {{Efficient
  iterative schemes for {\it ab initio} total-energy calculations using a
  plane-wave basis set}},\ }}\href@noop {} {\bibfield  {journal} {\bibinfo
  {journal} {Phys. Rev. B}\ }\textbf {\bibinfo {volume} {54}},\ \bibinfo
  {pages} {11169} (\bibinfo {year} {1996})}\BibitemShut {NoStop}%
\bibitem [{\citenamefont {Kresse}\ and\ \citenamefont {Joubert}(1999)}]{VASP2}%
  \BibitemOpen
  \bibfield  {author} {\bibinfo {author} {\bibfnamefont {G.}~\bibnamefont
  {Kresse}}\ and\ \bibinfo {author} {\bibfnamefont {D.}~\bibnamefont
  {Joubert}},\ }\bibfield  {title} {\emph {\bibinfo {title} {{From ultrasoft
  pseudopotentials to the projector augmented-wave method}},\ }}\href@noop {}
  {\bibfield  {journal} {\bibinfo  {journal} {Phys. Rev. B}\ }\textbf {\bibinfo
  {volume} {59}},\ \bibinfo {pages} {1758} (\bibinfo {year}
  {1999})}\BibitemShut {NoStop}%
\bibitem [{\citenamefont {Bl\"ochl}(1994)}]{PAW}%
  \BibitemOpen
  \bibfield  {author} {\bibinfo {author} {\bibfnamefont {P.~E.}\ \bibnamefont
  {Bl\"ochl}},\ }\bibfield  {title} {\emph {\bibinfo {title} {{Projector
  augmented-wave method}},\ }}\href@noop {} {\bibfield  {journal} {\bibinfo
  {journal} {Phys. Rev. B}\ }\textbf {\bibinfo {volume} {50}},\ \bibinfo
  {pages} {17953} (\bibinfo {year} {1994})}\BibitemShut {NoStop}%
\bibitem [{\citenamefont {Perdew}\ \emph {et~al.}(1996)\citenamefont {Perdew},
  \citenamefont {Burke},\ and\ \citenamefont {Ernzerhof}}]{PBE}%
  \BibitemOpen
  \bibfield  {author} {\bibinfo {author} {\bibfnamefont {J.~P.}\ \bibnamefont
  {Perdew}}, \bibinfo {author} {\bibfnamefont {K.}~\bibnamefont {Burke}}, \
  and\ \bibinfo {author} {\bibfnamefont {M.}~\bibnamefont {Ernzerhof}},\
  }\bibfield  {title} {\emph {\bibinfo {title} {{Generalized Gradient
  Approximation Made Simple}},\ }}\href@noop {} {\bibfield  {journal} {\bibinfo
   {journal} {Phys. Rev. Lett.}\ }\textbf {\bibinfo {volume} {77}},\ \bibinfo
  {pages} {3865} (\bibinfo {year} {1996})}\BibitemShut {NoStop}%
\bibitem [{\citenamefont {Perdew}\ \emph {et~al.}(2008)\citenamefont {Perdew},
  \citenamefont {Ruzsinszky}, \citenamefont {Csonka}, \citenamefont {Vydrov},
  \citenamefont {Scuseria}, \citenamefont {Constantin}, \citenamefont {Zhou},\
  and\ \citenamefont {Burke}}]{PBEsol}%
  \BibitemOpen
  \bibfield  {author} {\bibinfo {author} {\bibfnamefont {J.~P.}\ \bibnamefont
  {Perdew}}, \bibinfo {author} {\bibfnamefont {A.}~\bibnamefont {Ruzsinszky}},
  \bibinfo {author} {\bibfnamefont {G.~I.}\ \bibnamefont {Csonka}}, \bibinfo
  {author} {\bibfnamefont {O.~A.}\ \bibnamefont {Vydrov}}, \bibinfo {author}
  {\bibfnamefont {G.~E.}\ \bibnamefont {Scuseria}}, \bibinfo {author}
  {\bibfnamefont {L.~A.}\ \bibnamefont {Constantin}}, \bibinfo {author}
  {\bibfnamefont {X.}~\bibnamefont {Zhou}}, \ and\ \bibinfo {author}
  {\bibfnamefont {K.}~\bibnamefont {Burke}},\ }\bibfield  {title} {\emph
  {\bibinfo {title} {Restoring the density-gradient expansion for exchange in
  solids and surfaces},\ }}\href@noop {} {\bibfield  {journal} {\bibinfo
  {journal} {Phys. Rev. Lett.}\ }\textbf {\bibinfo {volume} {100}},\ \bibinfo
  {pages} {136406} (\bibinfo {year} {2008})}\BibitemShut {NoStop}%
\bibitem [{\citenamefont {Grimme}(2006)}]{VDW1}%
  \BibitemOpen
  \bibfield  {author} {\bibinfo {author} {\bibfnamefont {S.}~\bibnamefont
  {Grimme}},\ }\bibfield  {title} {\emph {\bibinfo {title} {Semiempirical
  gga-type density functional constructed with a long-range dispersion
  correction},\ }}\href@noop {} {\bibfield  {journal} {\bibinfo  {journal} {J.
  Comp. Chem.}\ }\textbf {\bibinfo {volume} {27}},\ \bibinfo {pages} {1787}
  (\bibinfo {year} {2006})}\BibitemShut {NoStop}%
\bibitem [{\citenamefont {Tkatchenko}\ and\ \citenamefont
  {Scheffler}(2009)}]{VDW2}%
  \BibitemOpen
  \bibfield  {author} {\bibinfo {author} {\bibfnamefont {A.}~\bibnamefont
  {Tkatchenko}}\ and\ \bibinfo {author} {\bibfnamefont {M.}~\bibnamefont
  {Scheffler}},\ }\bibfield  {title} {\emph {\bibinfo {title} {Accurate
  molecular van der waals interactions from ground-state electron density and
  free-atom reference data},\ }}\href@noop {} {\bibfield  {journal} {\bibinfo
  {journal} {Phys. Rev. Lett.}\ }\textbf {\bibinfo {volume} {102}},\ \bibinfo
  {pages} {073005} (\bibinfo {year} {2009})}\BibitemShut {NoStop}%
\bibitem [{\citenamefont {Klime\v{s}}\ \emph {et~al.}(2010)\citenamefont
  {Klime\v{s}}, \citenamefont {Bowler},\ and\ \citenamefont
  {Michaelides}}]{VDWopt}%
  \BibitemOpen
  \bibfield  {author} {\bibinfo {author} {\bibfnamefont {J.}~\bibnamefont
  {Klime\v{s}}}, \bibinfo {author} {\bibfnamefont {D.~R.}\ \bibnamefont
  {Bowler}}, \ and\ \bibinfo {author} {\bibfnamefont {A.}~\bibnamefont
  {Michaelides}},\ }\bibfield  {title} {\emph {\bibinfo {title} {Chemical
  accuracy for the van der waals density functional},\ }}\href@noop {}
  {\bibfield  {journal} {\bibinfo  {journal} {J. Phys.: Cond. Matt.}\ }\textbf
  {\bibinfo {volume} {22}},\ \bibinfo {pages} {022201} (\bibinfo {year}
  {2010})}\BibitemShut {NoStop}%
\bibitem [{\citenamefont {Steiner}\ \emph {et~al.}(2016)\citenamefont
  {Steiner}, \citenamefont {Khmelevskyi}, \citenamefont {Marsmann},\ and\
  \citenamefont {Kresse}}]{SOC_VASP}%
  \BibitemOpen
  \bibfield  {author} {\bibinfo {author} {\bibfnamefont {S.}~\bibnamefont
  {Steiner}}, \bibinfo {author} {\bibfnamefont {S.}~\bibnamefont
  {Khmelevskyi}}, \bibinfo {author} {\bibfnamefont {M.}~\bibnamefont
  {Marsmann}}, \ and\ \bibinfo {author} {\bibfnamefont {G.}~\bibnamefont
  {Kresse}},\ }\bibfield  {title} {\emph {\bibinfo {title} {{Calculation of the
  magnetic anisotropy with projected-augmented-wave methodology and the case
  study of disordered ${\mathrm{Fe}}_{1\ensuremath{-}x}{\mathrm{Co}}_{x}$
  alloys}},\ }}\href@noop {} {\bibfield  {journal} {\bibinfo  {journal} {Phys.
  Rev. B}\ }\textbf {\bibinfo {volume} {93}},\ \bibinfo {pages} {224425}
  (\bibinfo {year} {2016})}\BibitemShut {NoStop}%
\bibitem [{\citenamefont {Mostofi}\ \emph {et~al.}(2008)\citenamefont
  {Mostofi}, \citenamefont {Yates}, \citenamefont {Lee}, \citenamefont {Souza},
  \citenamefont {Vanderbilt},\ and\ \citenamefont {Marzari}}]{WANNIER90}%
  \BibitemOpen
  \bibfield  {author} {\bibinfo {author} {\bibfnamefont {A.~A.}\ \bibnamefont
  {Mostofi}}, \bibinfo {author} {\bibfnamefont {J.~R.}\ \bibnamefont {Yates}},
  \bibinfo {author} {\bibfnamefont {Y.-S.}\ \bibnamefont {Lee}}, \bibinfo
  {author} {\bibfnamefont {I.}~\bibnamefont {Souza}}, \bibinfo {author}
  {\bibfnamefont {D.}~\bibnamefont {Vanderbilt}}, \ and\ \bibinfo {author}
  {\bibfnamefont {N.}~\bibnamefont {Marzari}},\ }\bibfield  {title} {\emph
  {\bibinfo {title} {{Wannier90: A Tool for Obtaining Maximally-Localised
  Wannier Functions}},\ }}\href@noop {} {\bibfield  {journal} {\bibinfo
  {journal} {Comput. Phys. Commun.}\ }\textbf {\bibinfo {volume} {178}},\
  \bibinfo {pages} {685} (\bibinfo {year} {2008})}\BibitemShut {NoStop}%
\bibitem [{\citenamefont {Soluyanov}\ and\ \citenamefont
  {Vanderbilt}(2011)}]{WCC}%
  \BibitemOpen
  \bibfield  {author} {\bibinfo {author} {\bibfnamefont {A.~A.}\ \bibnamefont
  {Soluyanov}}\ and\ \bibinfo {author} {\bibfnamefont {D.}~\bibnamefont
  {Vanderbilt}},\ }\bibfield  {title} {\emph {\bibinfo {title} {{Computing
  topological invariants without inversion symmetry}},\ }}\href@noop {}
  {\bibfield  {journal} {\bibinfo  {journal} {Phys. Rev. B}\ }\textbf {\bibinfo
  {volume} {83}},\ \bibinfo {pages} {235401} (\bibinfo {year}
  {2011})}\BibitemShut {NoStop}%
\bibitem [{\citenamefont {Gresch}\ \emph {et~al.}(2017)\citenamefont {Gresch},
  \citenamefont {Aut\`es}, \citenamefont {Yazyev}, \citenamefont {Troyer},
  \citenamefont {Vanderbilt}, \citenamefont {Bernevig},\ and\ \citenamefont
  {Soluyanov}}]{z2pack}%
  \BibitemOpen
  \bibfield  {author} {\bibinfo {author} {\bibfnamefont {D.}~\bibnamefont
  {Gresch}}, \bibinfo {author} {\bibfnamefont {G.}~\bibnamefont {Aut\`es}},
  \bibinfo {author} {\bibfnamefont {O.~V.}\ \bibnamefont {Yazyev}}, \bibinfo
  {author} {\bibfnamefont {M.}~\bibnamefont {Troyer}}, \bibinfo {author}
  {\bibfnamefont {D.}~\bibnamefont {Vanderbilt}}, \bibinfo {author}
  {\bibfnamefont {B.~A.}\ \bibnamefont {Bernevig}}, \ and\ \bibinfo {author}
  {\bibfnamefont {A.~A.}\ \bibnamefont {Soluyanov}},\ }\bibfield  {title}
  {\emph {\bibinfo {title} {Z2pack: Numerical implementation of hybrid wannier
  centers for identifying topological materials},\ }}\href@noop {} {\bibfield
  {journal} {\bibinfo  {journal} {Phys. Rev. B}\ }\textbf {\bibinfo {volume}
  {95}},\ \bibinfo {pages} {075146} (\bibinfo {year} {2017})}\BibitemShut
  {NoStop}%
\bibitem [{\citenamefont {Parlinski}\ \emph {et~al.}(1997)\citenamefont
  {Parlinski}, \citenamefont {Li},\ and\ \citenamefont {Kawazoe}}]{Parlinski}%
  \BibitemOpen
  \bibfield  {author} {\bibinfo {author} {\bibfnamefont {K.}~\bibnamefont
  {Parlinski}}, \bibinfo {author} {\bibfnamefont {Z.~Q.}\ \bibnamefont {Li}}, \
  and\ \bibinfo {author} {\bibfnamefont {Y.}~\bibnamefont {Kawazoe}},\
  }\bibfield  {title} {\emph {\bibinfo {title} {First-principles determination
  of the soft mode in cubic ${Zr}{O}_{2}$},\ }}\href@noop {} {\bibfield
  {journal} {\bibinfo  {journal} {Phys. Rev. Lett.}\ }\textbf {\bibinfo
  {volume} {78}},\ \bibinfo {pages} {4063} (\bibinfo {year}
  {1997})}\BibitemShut {NoStop}%
\bibitem [{\citenamefont {Togo}\ \emph {et~al.}(2008)\citenamefont {Togo},
  \citenamefont {Oba},\ and\ \citenamefont {Tanaka}}]{PHONOPY}%
  \BibitemOpen
  \bibfield  {author} {\bibinfo {author} {\bibfnamefont {A.}~\bibnamefont
  {Togo}}, \bibinfo {author} {\bibfnamefont {F.}~\bibnamefont {Oba}}, \ and\
  \bibinfo {author} {\bibfnamefont {I.}~\bibnamefont {Tanaka}},\ }\bibfield
  {title} {\emph {\bibinfo {title} {First-principles calculations of the
  ferroelastic transition between rutile-type and ${Ca}{Cl}_{2}$-type
  ${Si}{O}_{2}$ at high pressures},\ }}\href@noop {} {\bibfield  {journal}
  {\bibinfo  {journal} {Phys. Rev. B}\ }\textbf {\bibinfo {volume} {78}},\
  \bibinfo {pages} {134106} (\bibinfo {year} {2008})}\BibitemShut {NoStop}%
\bibitem [{\citenamefont {Ku}\ \emph {et~al.}(2010)\citenamefont {Ku},
  \citenamefont {Berlijn},\ and\ \citenamefont {Lee}}]{WeiKu}%
  \BibitemOpen
  \bibfield  {author} {\bibinfo {author} {\bibfnamefont {W.}~\bibnamefont
  {Ku}}, \bibinfo {author} {\bibfnamefont {T.}~\bibnamefont {Berlijn}}, \ and\
  \bibinfo {author} {\bibfnamefont {C.-C.}\ \bibnamefont {Lee}},\ }\bibfield
  {title} {\emph {\bibinfo {title} {{Unfolding First-Principles Band
  Structures}},\ }}\href@noop {} {\bibfield  {journal} {\bibinfo  {journal}
  {Phys. Rev. Lett.}\ }\textbf {\bibinfo {volume} {104}},\ \bibinfo {pages}
  {216401} (\bibinfo {year} {2010})}\BibitemShut {NoStop}%
\bibitem [{\citenamefont {Tomi\ifmmode~\acute{c}\else \'{c}\fi{}}\ \emph
  {et~al.}(2014)\citenamefont {Tomi\ifmmode~\acute{c}\else \'{c}\fi{}},
  \citenamefont {Jeschke},\ and\ \citenamefont {Valent\'{\i}}}]{vaspunfold}%
  \BibitemOpen
  \bibfield  {author} {\bibinfo {author} {\bibfnamefont {M.}~\bibnamefont
  {Tomi\ifmmode~\acute{c}\else \'{c}\fi{}}}, \bibinfo {author} {\bibfnamefont
  {H.~O.}\ \bibnamefont {Jeschke}}, \ and\ \bibinfo {author} {\bibfnamefont
  {R.}~\bibnamefont {Valent\'{\i}}},\ }\bibfield  {title} {\emph {\bibinfo
  {title} {{Unfolding of electronic structure through induced representations
  of space groups: Application to Fe-based superconductors}},\ }}\href@noop {}
  {\bibfield  {journal} {\bibinfo  {journal} {Phys. Rev. B}\ }\textbf {\bibinfo
  {volume} {90}},\ \bibinfo {pages} {195121} (\bibinfo {year}
  {2014})}\BibitemShut {NoStop}%
\bibitem [{\citenamefont {Lander}\ and\ \citenamefont
  {Morrison}(1964)}]{Lander}%
  \BibitemOpen
  \bibfield  {author} {\bibinfo {author} {\bibfnamefont {J.}~\bibnamefont
  {Lander}}\ and\ \bibinfo {author} {\bibfnamefont {J.}~\bibnamefont
  {Morrison}},\ }\bibfield  {title} {\emph {\bibinfo {title} {{Surface
  reactions of silicon with aluminum and with indium}},\ }}\href@noop {}
  {\bibfield  {journal} {\bibinfo  {journal} {Surf. Sci.}\ }\textbf {\bibinfo
  {volume} {2}},\ \bibinfo {pages} {553} (\bibinfo {year} {1964})}\BibitemShut
  {NoStop}%
\bibitem [{\citenamefont {Northrup}(1984)}]{Northrup}%
  \BibitemOpen
  \bibfield  {author} {\bibinfo {author} {\bibfnamefont {J.~E.}\ \bibnamefont
  {Northrup}},\ }\bibfield  {title} {\emph {\bibinfo {title}
  {{$\mathrm{Si}(111)\sqrt{3}\ifmmode\times\else\texttimes\fi{}\sqrt{3}$-Al: An
  Adatom-Induced Reconstruction}},\ }}\href@noop {} {\bibfield  {journal}
  {\bibinfo  {journal} {Phys. Rev. Lett.}\ }\textbf {\bibinfo {volume} {53}},\
  \bibinfo {pages} {683} (\bibinfo {year} {1984})}\BibitemShut {NoStop}%
\bibitem [{\citenamefont {Nicholls}\ \emph {et~al.}(1987)\citenamefont
  {Nicholls}, \citenamefont {Reihl},\ and\ \citenamefont
  {Northrup}}]{Nicholls}%
  \BibitemOpen
  \bibfield  {author} {\bibinfo {author} {\bibfnamefont {J.~M.}\ \bibnamefont
  {Nicholls}}, \bibinfo {author} {\bibfnamefont {B.}~\bibnamefont {Reihl}}, \
  and\ \bibinfo {author} {\bibfnamefont {J.~E.}\ \bibnamefont {Northrup}},\
  }\bibfield  {title} {\emph {\bibinfo {title} {{Unoccupied surface states
  revealing the Si(111)\ensuremath{\surd}3 \ensuremath{\surd}3 -Al, -Ga, and
  -In adatom geometries}},\ }}\href@noop {} {\bibfield  {journal} {\bibinfo
  {journal} {Phys. Rev. B}\ }\textbf {\bibinfo {volume} {35}},\ \bibinfo
  {pages} {4137} (\bibinfo {year} {1987})}\BibitemShut {NoStop}%
\bibitem [{\citenamefont {Hamers}(1989)}]{Hamers}%
  \BibitemOpen
  \bibfield  {author} {\bibinfo {author} {\bibfnamefont {R.~J.}\ \bibnamefont
  {Hamers}},\ }\bibfield  {title} {\emph {\bibinfo {title} {{Effects of
  coverage on the geometry and electronic structure of ${Al}$ overlayers on
  ${Si}$(111)}},\ }}\href@noop {} {\bibfield  {journal} {\bibinfo  {journal}
  {Phys. Rev. B}\ }\textbf {\bibinfo {volume} {40}},\ \bibinfo {pages} {1657}
  (\bibinfo {year} {1989})}\BibitemShut {NoStop}%
\bibitem [{\citenamefont {Xiao}\ \emph {et~al.}(2007)\citenamefont {Xiao},
  \citenamefont {Yao},\ and\ \citenamefont {Niu}}]{GrapheneValley}%
  \BibitemOpen
  \bibfield  {author} {\bibinfo {author} {\bibfnamefont {D.}~\bibnamefont
  {Xiao}}, \bibinfo {author} {\bibfnamefont {W.}~\bibnamefont {Yao}}, \ and\
  \bibinfo {author} {\bibfnamefont {Q.}~\bibnamefont {Niu}},\ }\bibfield
  {title} {\emph {\bibinfo {title} {{Valley-Contrasting Physics in Graphene:
  Magnetic Moment and Topological Transport}},\ }}\href@noop {} {\bibfield
  {journal} {\bibinfo  {journal} {Phys. Rev. Lett.}\ }\textbf {\bibinfo
  {volume} {99}},\ \bibinfo {pages} {236809} (\bibinfo {year}
  {2007})}\BibitemShut {NoStop}%
\bibitem [{\citenamefont {Xiao}\ \emph {et~al.}(2012)\citenamefont {Xiao},
  \citenamefont {Liu}, \citenamefont {Feng}, \citenamefont {Xu},\ and\
  \citenamefont {Yao}}]{MoS2Valley}%
  \BibitemOpen
  \bibfield  {author} {\bibinfo {author} {\bibfnamefont {D.}~\bibnamefont
  {Xiao}}, \bibinfo {author} {\bibfnamefont {G.-B.}\ \bibnamefont {Liu}},
  \bibinfo {author} {\bibfnamefont {W.}~\bibnamefont {Feng}}, \bibinfo {author}
  {\bibfnamefont {X.}~\bibnamefont {Xu}}, \ and\ \bibinfo {author}
  {\bibfnamefont {W.}~\bibnamefont {Yao}},\ }\bibfield  {title} {\emph
  {\bibinfo {title} {{Coupled Spin and Valley Physics in Monolayers of
  ${\mathrm{MoS}}_{2}$ and Other Group-VI Dichalcogenides}},\ }}\href@noop {}
  {\bibfield  {journal} {\bibinfo  {journal} {Phys. Rev. Lett.}\ }\textbf
  {\bibinfo {volume} {108}},\ \bibinfo {pages} {196802} (\bibinfo {year}
  {2012})}\BibitemShut {NoStop}%
\bibitem [{\citenamefont {Liu}\ \emph {et~al.}(2011)\citenamefont {Liu},
  \citenamefont {Jiang},\ and\ \citenamefont {Yao}}]{Silicine}%
  \BibitemOpen
  \bibfield  {author} {\bibinfo {author} {\bibfnamefont {C.-C.}\ \bibnamefont
  {Liu}}, \bibinfo {author} {\bibfnamefont {H.}~\bibnamefont {Jiang}}, \ and\
  \bibinfo {author} {\bibfnamefont {Y.}~\bibnamefont {Yao}},\ }\bibfield
  {title} {\emph {\bibinfo {title} {{Low-energy effective Hamiltonian involving
  spin-orbit coupling in silicene and two-dimensional germanium and tin}},\
  }}\href@noop {} {\bibfield  {journal} {\bibinfo  {journal} {Phys. Rev. B}\
  }\textbf {\bibinfo {volume} {84}},\ \bibinfo {pages} {195430} (\bibinfo
  {year} {2011})}\BibitemShut {NoStop}%
\bibitem [{\citenamefont {Song}\ \emph {et~al.}(2018)\citenamefont {Song},
  \citenamefont {Soriano}, \citenamefont {Cummings}, \citenamefont {Robles},
  \citenamefont {Ordejón},\ and\ \citenamefont {Roche}}]{Roche}%
  \BibitemOpen
  \bibfield  {author} {\bibinfo {author} {\bibfnamefont {K.}~\bibnamefont
  {Song}}, \bibinfo {author} {\bibfnamefont {D.}~\bibnamefont {Soriano}},
  \bibinfo {author} {\bibfnamefont {A.~W.}\ \bibnamefont {Cummings}}, \bibinfo
  {author} {\bibfnamefont {R.}~\bibnamefont {Robles}}, \bibinfo {author}
  {\bibfnamefont {P.}~\bibnamefont {Ordejón}}, \ and\ \bibinfo {author}
  {\bibfnamefont {S.}~\bibnamefont {Roche}},\ }\bibfield  {title} {\emph
  {\bibinfo {title} {Spin proximity effects in graphene/topological insulator
  heterostructures},\ }}\href@noop {} {\bibfield  {journal} {\bibinfo
  {journal} {Nano Letters}\ }\textbf {\bibinfo {volume} {18}},\ \bibinfo
  {pages} {2033} (\bibinfo {year} {2018})}\BibitemShut {NoStop}%
\bibitem [{\citenamefont {Tang}\ \emph {et~al.}(2014)\citenamefont {Tang},
  \citenamefont {Chen}, \citenamefont {Cao}, \citenamefont {Huang},
  \citenamefont {Cahangirov}, \citenamefont {Xian}, \citenamefont {Xu},
  \citenamefont {Zhang}, \citenamefont {Duan},\ and\ \citenamefont
  {Rubio}}]{Tang}%
  \BibitemOpen
  \bibfield  {author} {\bibinfo {author} {\bibfnamefont {P.}~\bibnamefont
  {Tang}}, \bibinfo {author} {\bibfnamefont {P.}~\bibnamefont {Chen}}, \bibinfo
  {author} {\bibfnamefont {W.}~\bibnamefont {Cao}}, \bibinfo {author}
  {\bibfnamefont {H.}~\bibnamefont {Huang}}, \bibinfo {author} {\bibfnamefont
  {S.}~\bibnamefont {Cahangirov}}, \bibinfo {author} {\bibfnamefont
  {L.}~\bibnamefont {Xian}}, \bibinfo {author} {\bibfnamefont {Y.}~\bibnamefont
  {Xu}}, \bibinfo {author} {\bibfnamefont {S.-C.}\ \bibnamefont {Zhang}},
  \bibinfo {author} {\bibfnamefont {W.}~\bibnamefont {Duan}}, \ and\ \bibinfo
  {author} {\bibfnamefont {A.}~\bibnamefont {Rubio}},\ }\bibfield  {title}
  {\emph {\bibinfo {title} {Stable two-dimensional dumbbell stanene: A quantum
  spin hall insulator},\ }}\href@noop {} {\bibfield  {journal} {\bibinfo
  {journal} {Phys. Rev. B}\ }\textbf {\bibinfo {volume} {90}},\ \bibinfo
  {pages} {121408} (\bibinfo {year} {2014})}\BibitemShut {NoStop}%
\bibitem [{\citenamefont {Liu}\ \emph {et~al.}(2016)\citenamefont {Liu},
  \citenamefont {Every},\ and\ \citenamefont {Tom\'anek}}]{LiuZA}%
  \BibitemOpen
  \bibfield  {author} {\bibinfo {author} {\bibfnamefont {D.}~\bibnamefont
  {Liu}}, \bibinfo {author} {\bibfnamefont {A.~G.}\ \bibnamefont {Every}}, \
  and\ \bibinfo {author} {\bibfnamefont {D.}~\bibnamefont {Tom\'anek}},\
  }\bibfield  {title} {\emph {\bibinfo {title} {Continuum approach for
  long-wavelength acoustic phonons in quasi-two-dimensional structures},\
  }}\href@noop {} {\bibfield  {journal} {\bibinfo  {journal} {Phys. Rev. B}\
  }\textbf {\bibinfo {volume} {94}},\ \bibinfo {pages} {165432} (\bibinfo
  {year} {2016})}\BibitemShut {NoStop}%
\bibitem [{\citenamefont {Peng}\ \emph {et~al.}(2016)\citenamefont {Peng},
  \citenamefont {Zhang}, \citenamefont {Shao}, \citenamefont {Xu},
  \citenamefont {Zhang},\ and\ \citenamefont {Zhu}}]{Peng}%
  \BibitemOpen
  \bibfield  {author} {\bibinfo {author} {\bibfnamefont {B.}~\bibnamefont
  {Peng}}, \bibinfo {author} {\bibfnamefont {H.}~\bibnamefont {Zhang}},
  \bibinfo {author} {\bibfnamefont {H.}~\bibnamefont {Shao}}, \bibinfo {author}
  {\bibfnamefont {Y.}~\bibnamefont {Xu}}, \bibinfo {author} {\bibfnamefont
  {R.}~\bibnamefont {Zhang}}, \ and\ \bibinfo {author} {\bibfnamefont
  {H.}~\bibnamefont {Zhu}},\ }\bibfield  {title} {\emph {\bibinfo {title} {The
  electronic{,} optical{,} and thermodynamic properties of borophene from
  first-principles calculations},\ }}\href@noop {} {\bibfield  {journal}
  {\bibinfo  {journal} {J. Mater. Chem. C}\ }\textbf {\bibinfo {volume} {4}},\
  \bibinfo {pages} {3592} (\bibinfo {year} {2016})}\BibitemShut {NoStop}%
\bibitem [{\citenamefont {Yu}\ \emph {et~al.}(2016)\citenamefont {Yu},
  \citenamefont {Niu}, \citenamefont {Zhu}, \citenamefont {Wang},\ and\
  \citenamefont {Zhang}}]{Yu}%
  \BibitemOpen
  \bibfield  {author} {\bibinfo {author} {\bibfnamefont {W.}~\bibnamefont
  {Yu}}, \bibinfo {author} {\bibfnamefont {C.-Y.}\ \bibnamefont {Niu}},
  \bibinfo {author} {\bibfnamefont {Z.}~\bibnamefont {Zhu}}, \bibinfo {author}
  {\bibfnamefont {X.}~\bibnamefont {Wang}}, \ and\ \bibinfo {author}
  {\bibfnamefont {W.-B.}\ \bibnamefont {Zhang}},\ }\bibfield  {title} {\emph
  {\bibinfo {title} {Atomically thin binary v–v compound semiconductor: a
  first-principles study},\ }}\href@noop {} {\bibfield  {journal} {\bibinfo
  {journal} {J. Mater. Chem. C}\ }\textbf {\bibinfo {volume} {4}},\ \bibinfo
  {pages} {6581} (\bibinfo {year} {2016})}\BibitemShut {NoStop}%
\bibitem [{\citenamefont {Rivero}\ \emph {et~al.}(2014)\citenamefont {Rivero},
  \citenamefont {Yan}, \citenamefont {Garc\'{\i}a-Su\'arez}, \citenamefont
  {Ferrer},\ and\ \citenamefont {Barraza-Lopez}}]{Rivero}%
  \BibitemOpen
  \bibfield  {author} {\bibinfo {author} {\bibfnamefont {P.}~\bibnamefont
  {Rivero}}, \bibinfo {author} {\bibfnamefont {J.-A.}\ \bibnamefont {Yan}},
  \bibinfo {author} {\bibfnamefont {V.~M.}\ \bibnamefont
  {Garc\'{\i}a-Su\'arez}}, \bibinfo {author} {\bibfnamefont {J.}~\bibnamefont
  {Ferrer}}, \ and\ \bibinfo {author} {\bibfnamefont {S.}~\bibnamefont
  {Barraza-Lopez}},\ }\bibfield  {title} {\emph {\bibinfo {title} {Stability
  and properties of high-buckled two-dimensional tin and lead},\ }}\href@noop
  {} {\bibfield  {journal} {\bibinfo  {journal} {Phys. Rev. B}\ }\textbf
  {\bibinfo {volume} {90}},\ \bibinfo {pages} {241408} (\bibinfo {year}
  {2014})}\BibitemShut {NoStop}%
\end{thebibliography}%

\end{document}